\documentclass[journal]{IEEEtran}[11pt]
\ifCLASSINFOpdf
\usepackage[pdftex]{graphicx}
\graphicspath{{../pdf/}{../jpeg/}}
\DeclareGraphicsExtensions{.pdf,.jpeg,.png}
\else
\usepackage[dvips]{graphicx}
\graphicspath{{../eps/}}
\DeclareGraphicsExtensions{.eps}
\fi\usepackage{graphicx}
\usepackage{subfloat}
\usepackage{graphics}
\usepackage{epsfig}
\usepackage{epstopdf}
\usepackage{stfloats}
\usepackage[cmex10]{amsmath}
\usepackage{algorithmic}
\usepackage{array}
\usepackage{mdwmath}
\usepackage{mdwtab}
\usepackage{graphicx}
\usepackage{subfigure}
\usepackage{color}
\usepackage{amsfonts,amssymb}
\usepackage{hyperref} %
\usepackage{multirow}
\usepackage{diagbox}
\usepackage{caption}
\usepackage{makecell}
\usepackage{lineno}
\usepackage{mathrsfs}
\usepackage{booktabs}
\usepackage{makecell}
\usepackage{tabu}
\usepackage{multirow}
\usepackage{multicol}
\usepackage{multirow}
\usepackage{float}
\usepackage{makecell}
\usepackage{booktabs}
\usepackage{amssymb}
\usepackage{bbding}
\usepackage[linesnumbered,ruled]{algorithm2e}
\usepackage{amsfonts,amsthm,array}
\usepackage[ruled]{algorithm2e}

\usepackage{threeparttable}

\begin{document}
	
	
\title{3D Trajectory Design for Energy-constrained Aerial CRNs Under Probabilistic LoS Channel \thanks{Manuscript received.}}

    \author{Hongjiang~Lei, 
    	Xiaqiu~Wu,
    	Ki-Hong~Park, 
    	and
    	Gaofeng~Pan
	\thanks{Hongjiang~Lei and Xiaqiu~Wu are with the School of Communications and Information Engineering, Chongqing University of Posts and Telecommunications, Chongqing 400065, China (e-mail: leihj@cqupt.edu.cn, cquptwxq@163.com).}
	\thanks{Ki-Hong~Park is with the CEMSE Division, King Abdullah University of Science and Technology (KAUST), Thuwal 23955-6900, Saudi Arabia (e-mail: kihong.park@kaust.edu.sa).}
	\thanks{Gaofeng~Pan is with the School of Cyberspace Science and Technology, Beijing Institute of Technology, Beijing 100081, China (e-mail: gaofeng.pan.cn@ieee.org).}
}

\maketitle
	
\begin{abstract}

Unmanned aerial vehicles (UAVs) have been attracting significant attention because there is a high probability of line-of-sight links being obtained between them and terrestrial nodes in high-rise urban areas. 
In this work, we investigate cognitive radio networks (CRNs) by jointly designing three-dimensional (3D) trajectory, the transmit power of the UAV, and user scheduling.
Considering the UAV's onboard energy consumption, an optimization problem is formulated in which the average achievable rate of the considered system is maximized by jointly optimizing the UAV's 3D trajectory, transmission power, and user scheduling. 
Due to the non-convex optimization problem, a lower bound on the average achievable rate is utilized to reduce the complexity of the solution. Subsequently, the original optimization problem is decoupled into four subproblems by using block coordinate descent, and each subproblem is transformed into manageable convex optimization problems by introducing slack variables and successive convex approximation. Numerical results validate the effectiveness of our proposed algorithm and demonstrate that the 3D trajectories of UAVs can enhance the average achievable rate of aerial CRNs.

\end{abstract}
	

\section{Introduction}
\label{sec:introduction}

\subsection{Background and Related Works}
\label{Background}

Unmanned aerial vehicles (UAVs) have become a revolutionary technology with extensive applications in various fields. UAVs were initially developed for military purposes and are now widely used in civilian and commercial sectors such as agriculture, disaster management, surveillance, and aerial photography, including surveillance and monitoring, aerial imaging, precision agriculture, intelligent logistics, law enforcement, disaster response, and prehospital emergency \cite{WuQ2021JSAC}. The rapid development of UAV technology, including the improvement of autonomy, payload capacity, and flight endurance, has promoted the widespread application of UAVs in various industries worldwide \cite{LiB2019IOT,ZengY2019P}. The continuous development of UAV technology has brought new opportunities and challenges, requiring further exploration and research in areas such as regulation, security, and integration into existing infrastructure. Modifying the positioning or mapping out the flight path of UAVs can ensure a stable line-of-sight (LoS) connection to terrestrial nodes (TNs) with optimal reliability. Consequently, the altitude and lateral placement of UAVs play a crucial role in enhancing the efficiency of UAV communication systems. The trajectory designing has emerged as a critical challenge to address in the development of UAV-assisted communication systems \cite{WuQ2019WC}.

To enhance the performance of the Internet of Thing (IoT) systems,  various schemes have been proposed in different scenarios. UAVs can play different roles in communication systems, such as serving as aerial base stations (BSs) to transmit information to ground users \cite{WuQ2018TWC}-\cite{CuiM2018TVT}.
In \cite{WuQ2018TWC}, a multi-purpose UAV-enabled wireless network was considered. Specifically, user scheduling, UAV trajectories, and transmission power were jointly optimized to maximize the minimum average rate among all users. 
The results indicated that, compared to traditional static base stations, the mobility of UAVs offers advantages in achieving better air-to-ground (A2G) channel conditions and providing additional flexibility for interference mitigation, thereby enhancing system throughput.
Research on secure UAV-assisted downlink transmission in the presence of colluding eavesdroppers was presented in \cite{LiZ2019CL}.
A single-antenna UAV served multiple ground users with imperfect location information, utilizing a power-splitting scheme to transmit confidential information and artificial noise.
In \cite{FuH2022TVT}, a scenario was considered where UAVs provide services to a group of TNs and maximize the minimum secrecy rate (SR) to ensure fairness between TNs. The results indicated that the minimum SR was significantly improved.
A UAV-to-ground communication system operating in the presence of multiple potential Eves was investigated in \cite{CuiM2018TVT} with incomplete information about the Eves' locations.  effectively.

Zeng \textit{et al.}  derived the closed-form expression of propulsion energy consumption for the fixed-wing UAVs in \cite{ZengY2017TWC}.
Their results demonstrated that UAVs must operate within specific speed and acceleration limits to conserve energy.
The propulsion energy consumption for the fixed-wing UAVs is a function of the flying velocity and acceleration.
Moreover, for level flight with fixed altitude, the UAV's energy consumption only depends on the velocity and acceleration rather than its actual location.
Considering the propulsion energy constraint, the authors in \cite{EomS2020TVT} investigated the minimum average rate maximization and the energy efficiency (EE) maximization problems by jointly designing the trajectory, velocity, and acceleration of the UAV and the transmit power of the TNs.
Considering the UAV's mobility constraints, the energy efficiency of the UAV-enabled communication system with multiple terrestrial jammer was maximized by designing the trajectory of the UAV in \cite{WuY2021WCL}.
The EE of the UAV communication systems was also considered in \cite{ZhangJ2020WCL}. To reduce the computational complexity, a new method based on receding horizon optimization was introduced to solve the formulated problem.
The authors in \cite{BejaouiA2020WCL} studied the max-min fairness problem of a system consisting of multiple UAVs and multiple TNs. The minimum achievable rate was maximized by jointly designing the UAVs' trajectories, power allocation, and user scheduling.

Rotary-wing UAVs have the advantage of being able to take off and land vertically, as well as hover in a stationary position.
Zeng \textit{et al.}  derived the closed-form expression of propulsion energy consumption models for the rotary-wing UAVs in \cite{ZengY2019TWC}.
Their results have shown that the propulsion power consumption of rotary-wing UAVs consists of three components: blade profile, induced, and parasite power.
The first two parts increase with increasing speed, and the third part decreases with increasing speed.
The authors in \cite{ZhanC2019WCL} minimized the maximum energy consumption of the rotary-wing UAV-enabled IoT system by jointly designing the UAV's trajectory, user scheduling, and power allocation.
Considering the average and peak power constraints, the authors in \cite{DuoB2020TVFD}  maximized the secrecy EE (SEE) of the full-duplex UAV communication systems by jointly designing the trajectory and the power allocation.
Their results show that the SEE gains depended on the capability of the self-interference cancelation at the full-duplex UAV.
In \cite{ZhangR2021TWC}, the worst-case average SR and secrecy energy efficiency of the dual-UAV communication system were maximized, respectively. Their results demonstrated there is a tradeoff between maximizing the total information bits and minimizing the total propulsion energy consumption.
The authors in \cite{LiM2022IOT} investigated the passive eavesdropping scenario without the eavesdropper's instantaneous channel state information. Considering the constraints of connection outage probability, secrecy outage probability, and securely collected bits, the SEE was maximized by jointly optimizing the user scheduling and transmit power, UAV's transmit power, trajectory, codeword rate, and redundancy rate.
All the previous works utilized the deterministic LoS channel model which is usually a valid assumption for the scenarios without high and dense obstacles and the two-dimensional (2D) UAV trajectory was designed with a fixed altitude.

When the UAV flies at a relatively low altitude, the shadowing effect becomes more significant because obstacles will occasionally block the signal propagation between the UAV and the TN.
The probabilistic LoS (PLoS) channel model was proposed in \cite{AlHouraniA2014WCL}, and their results demonstrated that the probability of the LoS are functions of the elevation angle.
Specifically, there are two states for the A2G channel, LoS and non-LoS (NLoS), and the probabilities of LoS/NLoS states depend on the relative position between the UAV and TN and the distributions of building density and height.
Intuitively, the LoS probability increases with the elevation angle, by either moving the UAV horizontally closer to the ground node or increasing its altitude. 
In \cite{DuoB2021ICC}, a UAV was utilized to transmit jamming signals to enhance the terrestrial communication link.
The expected SR was maximized by jointly optimizing the transmit power at the BS, the jamming power of the UAV, and the 2D trajectory of the UAV.
In \cite{YouC2020TWC}, the minimum average data collection rate was maximized by jointly optimizing the three-dimensional (3D) trajectory, the flying speed, and user scheduling.
In \cite{DuoB2020TVT3D}, the authors considered an aerial wireless sensor network with a malicious terrestrial jammer and maximized the minimum expected rate by jointly optimizing the transmission scheduling and the 3D trajectory.
In \cite{GuanZ2021ICCC}, a UAV was utilized as a relay to transmit signals to the TN; the EE was maximized by optimizing the 3D trajectory of the UAV considering the energy consumption of the UAV.
In \cite{MengA2022IoT}, the authors considered the fairness among the TNs and maximized the minimum expected sum throughput of the TNs by jointly designing the 3D trajectory of the UAV and the TN scheduling.
In \cite{LeiH2024TCOM}, the authors studied the problem of secure data collection problem in aerial communication systems with multiple
location-uncertain terrestrial eavesdroppers.
The bandwidth allocation and the 3D trajectory of the UAV were jointly designed to maximize the system's overall fair SR subject to flight energy consumption, user fairness, and secure transmission constraints.

Cognitive radio technology is considered an advanced technology that can solve the problem of spectrum scarcity in UAV-assisted communication systems by using dynamic spectrum access techniques \cite{SaleemY2015JNCA}.
In \cite{LeiH2023IoT}, the authors considered the aerial underlay IoT systems with a single location-uncertainty eavesdropper.  
The cognitive UAV's 2D trajectory, transmit power, and user scheduling were jointly designed to maximize the average SR of the cognitive users (CUs).
The authors in \cite{LeiH2024TVT} considered the aerial  IoT systems with multiple location-uncertainty  full-duplex eavesdroppers worked in colluding mode, and the cognitive UAV's 2D trajectory and transmit power were jointly designed to maximize the worst average SR of the CUs.
In \cite{NguyenX2021TVT}, a UAV was utilized as a friendly jammer to transmit artificial noise for interrupting the eavesdropper in both scenarios with perfect and imperfect locations of the eavesdropper and TN. The average SR of the CU was maximized by jointly optimizing the transmit power and UAV's 3D trajectory.
In \cite{WangZ2021CC}, a UAV was utilized as a cognitive relay to forward the signals to CUs and the sum throughput of the CUs was maximized by jointly optimizing the power allocation and 3D trajectory.
It should be noted that LoS channel model was utilized in \cite{NguyenX2021TVT}  and \cite{WangZ2021CC}, although the 3D trajectory was designed.
Based on the PLoS channel model, the authors in \cite{HuangY2019TCOM} considered both the quasi-stationary UAV scenario and mobile UAV scenario. The CU's achievable rate was maximized by jointly designing the UAV's 3D trajectory and power control, subject to the UAV's altitude and power constraints and the interference temperature  (IT) constraint for the primary user (PU).
In  \cite{JiangY2021CC}, the SR of the UAV-enabled cognitive radio network (CRN) was maximized by optimizing UAV's 3D trajectory, velocity, and acceleration considering the requirements location constraint, the speed constraint of UAV, and the IT constraint for all the PU.

\begin{table*}
	\centering
	{
		\caption{ \textit{Related Works to Trajectory Design of UAV Communication Systems}.}
		\label{table1}
		\begin{threeparttable}
			\begin{tabular}{c|c|p{1.3cm}<{\centering}|c|c|c|c|c|c}
				\Xhline{1.2pt}
				\textbf{Reference} & \textbf{Role of UAVs} & \textbf{LoS/PLoS} & \textbf{\makecell[c]{2D/3D \\trajectory}} & \textbf{\makecell[c]{Propulsion \\ Limitation}}& \textbf{Rotary/Fixed} & \textbf{\makecell[c]{Power\\Control}} & \textbf{\makecell[c]{Transmission \\Scheduling}} & \textbf{CRNs}\\
				\hline
				\cite{WuQ2018TWC}  & Transmitter   & LoS         & 2D &              &              & $\checkmark$ & $\checkmark$  & \\
				\hline
				\cite{LiZ2019CL}  & Transmitter   & LoS         & 2D &              &              & $\checkmark$ & $\checkmark$  & \\
				\hline
				\cite{FuH2022TVT}  & Transmitter   & LoS         & 2D &              &              & $\checkmark$ & $\checkmark$  & \\
                \hline
				\cite{CuiM2018TVT}  & Transmitter   & LoS         & 2D &              &              & $\checkmark$ & $\checkmark$  & \\
				\hline

				\cite{ZengY2017TWC}  & Receiver  & LoS         & 2D & $\checkmark$   &     Fixed       &  &   & \\
				\hline
				\cite{EomS2020TVT}  & Receiver   & LoS         & 2D & $\checkmark$   &      Fixed    & $\checkmark$ & $\checkmark$  & \\
				\hline
				\cite{WuY2021WCL}  & Receiver   & LoS         & 2D &  $\checkmark$   &    Fixed     &  &   & \\
                \hline
				\cite{ZhangJ2020WCL}  & Receiver   & LoS         & 2D &  $\checkmark$    &    Fixed    &  & $\checkmark$  & \\
				\hline
				\cite{BejaouiA2020WCL}  & Transmitter   & LoS         & 2D &    $\checkmark$      &     Fixed     & $\checkmark$ & $\checkmark$  & \\
				\hline

				\cite{ZengY2019TWC}  & Receiver  & LoS         & 2D & $\checkmark$   &     Rotary      &  & $\checkmark$  & \\
				\hline
				\cite{ZhanC2019WCL}  & Receiver   & LoS         & 2D & $\checkmark$   &      Rotary   & $\checkmark$ & $\checkmark$  & \\
				\hline
				\cite{DuoB2020TVFD}  & Receiver   & LoS         & 2D &  $\checkmark$   &    Rotary     & $\checkmark$ &   & \\
                \hline
				\cite{ZhangR2021TWC}  & Receiver and Jammer  & LoS   & 2D &  $\checkmark$    &   Rotary  & $\checkmark$ & $\checkmark$  & \\
				\hline
				\cite{LiM2022IOT}  & Receiver  & LoS  & 2D &    $\checkmark$      &     Rotary    & $\checkmark$ & $\checkmark$  & \\
				\hline


				\cite{DuoB2021ICC}  & Jammer & PLoS   & 2D &   &          & $\checkmark$ &   & \\
				\hline
				\cite{YouC2020TWC}  & Receiver  & PLoS       & 3D &   &          &  & $\checkmark$  & \\
				\hline
				\cite{DuoB2020TVT3D}  & Receiver   & PLoS    & 3D &   &        &  & $\checkmark$  & \\
				\hline
				\cite{GuanZ2021ICCC}  & Relay  & PLoS       & 3D &   &          &  &   & \\
				\hline
				\cite{MengA2022IoT}  & Receiver   & PLoS     & 3D &  $\checkmark$   &    Rotary     &  & $\checkmark$  & \\
                \hline
				\cite{LeiH2024TCOM}  & Transmitter  & PLoS  & 3D &    $\checkmark$      &     Rotary    & $\checkmark$ & $\checkmark$  & \\
				\hline

				\cite{LeiH2023IoT}  & Receiver and Jammer & LoS       & 2D &   &          & $\checkmark$ & $\checkmark$  & $\checkmark$\\
				\hline
				\cite{LeiH2024TVT}  & Transmitter   & LoS    & 2D &   &        & $\checkmark$ & $\checkmark$  & $\checkmark$ \\
				\hline
				\cite{NguyenX2021TVT}  & Jammer   & LoS   & 3D &    &      & $\checkmark$ &  & $\checkmark$\\
                \hline
				\cite{WangZ2021CC}  & Relay & LoS  & 3D &         &        & $\checkmark$ & $\checkmark$  & $\checkmark$ \\
				\hline
				\cite{HuangY2019TCOM}  & Cognitive Transmitter  & LoS     & 3D &    &       &$\checkmark$  &  $\checkmark$ & $\checkmark$\\
                \hline
				\cite{JiangY2021CC}  & Cognitive Transmitter  & PLoS  & 3D &       &      & $\checkmark$ & $\checkmark$ & $\checkmark$\\
				\hline

				Our work   & Transmitter &  PLoS  & 3D  & $\checkmark$ & Rotary  & $\checkmark$   & $\checkmark$ & $\checkmark$   \\
				\Xhline{1.2pt}
			\end{tabular}
	\end{threeparttable}}
\end{table*}

\subsection{Motivation and Contributions}
\label{Motivation}

The discussed works proved that the performance of aerial CRNs was significantly improved by designing the UAV's trajectory and transmit power. 
However, these outstanding works on aerial CRNs did not consider the propulsion energy constraint and PLoS channel model in designing the cognitive UAV's trajectory.
In this work, based on the PLoS channel model, we consider the limited onboard energy and jointly design the cognitive UAV's 3D trajectory and transmission power and user scheduling to maximize the performance of the underlay IoT systems.
The main contributions of this work are summarized as follows:

\begin{enumerate}
	
	\item We consider an underlay aerial IoT system with multiple terrestrial cognitive users and a primary user on the ground. Based on the PLoS channel model, considering the UAV's onboard energy consumption, the average achievable rate of the considered system is maximized by jointly optimizing the UAV's 3D trajectory, transmission power, and user scheduling.
	The lower bound on the average achievable rate is utilized to reduce the complexity of the solution and the original optimization problem is decoupled into four subproblems by using the block coordinate descent (BCD). Each subproblem is transformed into manageable convex optimization problems by introducing slack variables and the successive convex approximation (SCA).

    \item The simulation results of the proposed scheme are compared with benchmark schemes, demonstrating that the proposed trajectory scheme effectively improves the average achievable rate of CRN IoT systems.
    Moreover, in CRNs, IT threshold, flight altitude, and UAV energy consumption are very crucial for trajectory design.

	\item Although the 3D trajectory design of UAVs was investigated in some outstanding works, such as \cite{YouC2020TWC} and \cite{DuoB2020TVT3D}, the energy consumption was ignored. Considering the energy consumption, IT threshold, and transmitting power over 3D trajectory design in this work makes the formulated optimization problem more challenging to solve.
	

   \item Relative to \cite{NguyenX2021TVT}-\cite{HuangY2019TCOM}, wherein the 3D trajectory design in the CRNs was investigated based on the LoS channel model, the vertical position of the UAV only depends on the IT constraint. In the CRNs based on the PLoS channel model, optimizing the vertical position of the UAV must consider both the IT constraints and the loss probability. Although the 3D trajectory design based on the PLoS model was considered in \cite{JiangY2021CC}, the elevation angle was assumed to be fixed to simplify the trajectory design. This work considers both the PLoS channel model and the energy consumption of the rotary-wing UAV in designing the 3D trajectory.


\end{enumerate}

\section{System Model}
\label{SystemModel}

\begin{figure}[t]
	\centering		
	\includegraphics[width = 2 in]{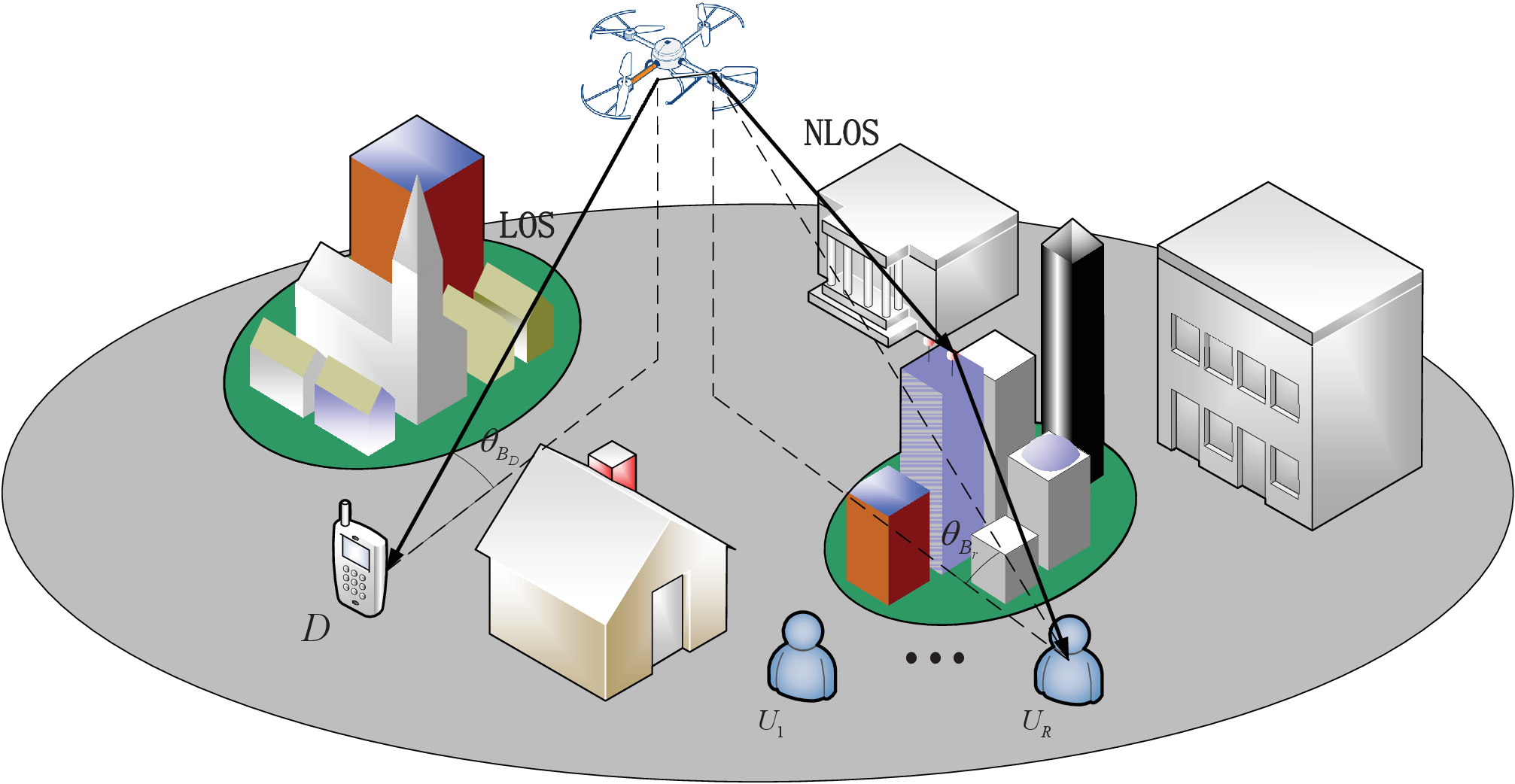}
    \caption{UAV-enabled CRN consists of an aerial base station (${B}$),  multiple cognitive users (${U_r}$), and a primary user (${D}$). The communication between UAV and TNs has corresponding LoS and NLoS probabilities based on their elevation angles.    }
    \label{fig1model}
\end{figure}

As shown in Fig. \ref{fig1model}, we consider an underlay  IoT system where an aerial base station $\left( B \right)$ transmits information to ground users.
There are multiple cognitive users $\left(U_r, r=1, \cdots, R\right)$ and a primary user $\left( D \right)$ on the ground. All the nodes are equipped with a single antenna.
The flight period of the UAV is $T$, and the period is decomposed into $N$ time slots.
The entire time is discredited as $n \in \mathcal {N} \triangleq \{1,\ldots,N\}$, where the time interval for each time slot is $\delta_t=\frac {T}{N}$. When $\delta_t$ is small enough, the position of the UAV can be considered stationary at each time slot \cite{WuQ2018TWC,LeiH2023IoT}.
In this work, a 3D Cartesian coordinate system is utilized and the horizontal positions of  $U_r $ and  $D$ are expressed as
$\mathbf{w}_{U_r}=\left[x_r, y_r\right]^T \in \mathbb{R}^{2 \times 1}$
and
$\mathbf{w}_D=\left[x_D, y_D\right]^T \in \mathbb{R}^{2 \times 1}$, respectively.

The PLoS model is utilized  and the LoS probability between $B$ and the ground node $X \left( {X \in \left\{ {{U_r},D} \right\}} \right)$  in the $n$th time slot  is expressed as
$	P_{BX}^{\mathrm{L}}\left[ n \right] = {\left( {1 + a{e^{\left( { - b\left[ {{\theta _{BX}}\left[ n \right] - a} \right]} \right)}}} \right)^{ - 1}}$,
where $a > 0$ and $b > 0$ are constants specified by the actual environment and the elevation angles between $B$ and $X$ \cite{AlHouraniA2014WCL} are expressed as
\begin{subequations}
	\begin{align}
		  {\theta _{B{U_r}}}\left[ n \right]&= \frac{{180}}{\pi }\arctan \left( {\frac{{{z_B}\left[ {n} \right]}}{{\parallel {{\mathbf{q}}_B}\left[ n \right]- {{\mathbf{w}}_{{U_r}}}\parallel }}} \right),\forall n,r, \label{eq1athetaBU} \\
		{\theta _{{BD}}}\left[ n \right]&= \frac{{180}}{\pi }\arctan \left( {\frac{{{z_B}\left[ {n} \right]}}{{\parallel {{\mathbf{q}}_B}\left[ n \right]- {{\mathbf{w}}_D}\parallel }}} \right),\forall n, \label{eq1bthetaBD}
	\end{align}
\end{subequations}
where 
${{\bf{q}}_B}$ and $z_B$ denote the horizontal and vertical position of  $B$, respectively.

The channel coefficient between $B$ and $X$ in the $n$th time slot is expressed as
\begin{align}\label{hBk}
	h_{BX}\left[ n \right]=\left\{\begin{array}{ll}
		\rho_0 d_{BX}\left[ n \right]^{-\alpha_L}, & \mathrm { LoS}, \\
		\mu\rho_0 d_{BX}\left[ n \right]^{-\alpha_N}, & \mathrm { NLoS},
	\end{array}\right.
\end{align}
where
$d_{BX}\left[ n \right]=\sqrt{z_B\left[ n \right]^2+\left\|\mathbf{q}_B\left[ n \right]-\mathbf{w}_{X}\right\|^2}$,
$\rho_0$ denotes the channel gain with the unit reference distance in the LoS environment,
$\mu $ is the additional signal attenuation factor for the NLoS environment, and $\alpha_L$ and $\alpha_N$ are the path loss exponents for LoS and NLoS environments, typically satisfying $\alpha_L$$\leq $$\alpha_N$ \cite{MengA2022IoT}.
Similar to \cite{DuoB2020TVT3D, LeiH2023IoT}, the small-scale fading is not considered in this work. Therefore, the expected rate of $U_r$ in the $n$th time slot is expressed as \cite{DuoB2021ICC}
\begin{align}\label{expectedR1}
	\bar {R}_{r}\left[ n \right]  = P_{BU_r}^{\mathrm{L}}\left[ n \right] R_r^{\mathrm{L}}\left[ n \right] + P_{BU_r}^{\mathrm{N}}\left[ n \right]R_r^{\mathrm{N}}\left[ n \right],
\end{align}
where
$R_r^{\mathrm{L}}\left[ n \right]=\log _2\left(1+\frac{P_B\left[ n \right]\gamma}{d_{BU_r}^{\alpha_{L}}\left[ n \right]}\right)$,
$R_r^{\mathrm{N}}\left[ n \right]=\log _2\left(1+\frac{P_B\left[ n \right]\gamma\mu}{d_{BU_r}^{\alpha_{N}}\left[ n \right]}\right)$,
$\gamma=\frac{\rho_0}{\sigma^2}$, where $P_B\left[ n \right]$ represents the transmission power of $B$, and $\sigma^2$ represents the variance of the additive white Gaussian noise (AWGN).
Because the rate in the NLoS scenario is much lower than that in the LoS scenario \cite{MengA2022IoT},
the average rate of the system is approximated as
\begin{align}\label{Rrlb}
	\bar{R}_{r}^{\mathrm{lb}}\left[ n \right]= P_{BU_r}^{\mathrm{L}}\left[ n \right] \log _2\left(1+\frac{P_B\left[ n \right]\gamma}{d_{BU_r}^{\alpha_{L}}\left[ n \right]}\right).
\end{align}

\begin{figure*}[t]
	\centering
	\subfigure[The different schemes.]{
		\label{fig02a}
		\includegraphics[width = 0.23  \textwidth]{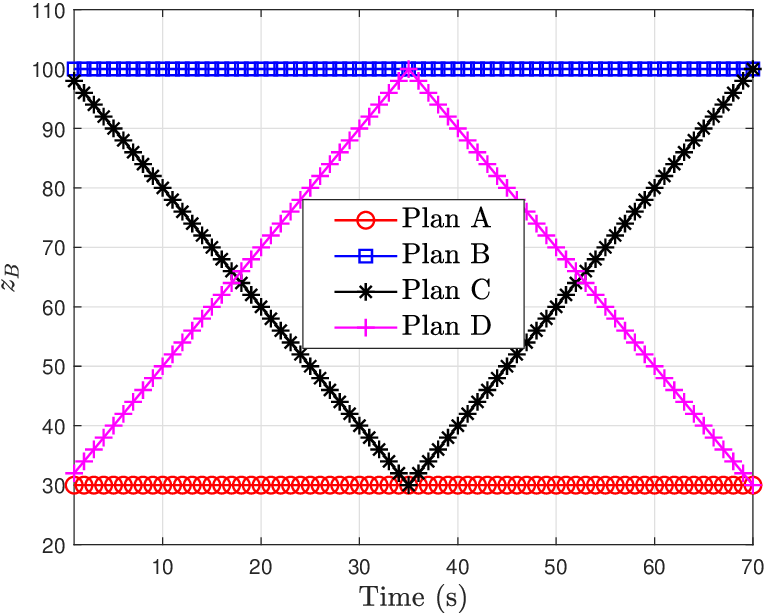}}
	\subfigure[The  LoS probability.]{
		\label{fig02b}
		\includegraphics[width = 0.23  \textwidth]{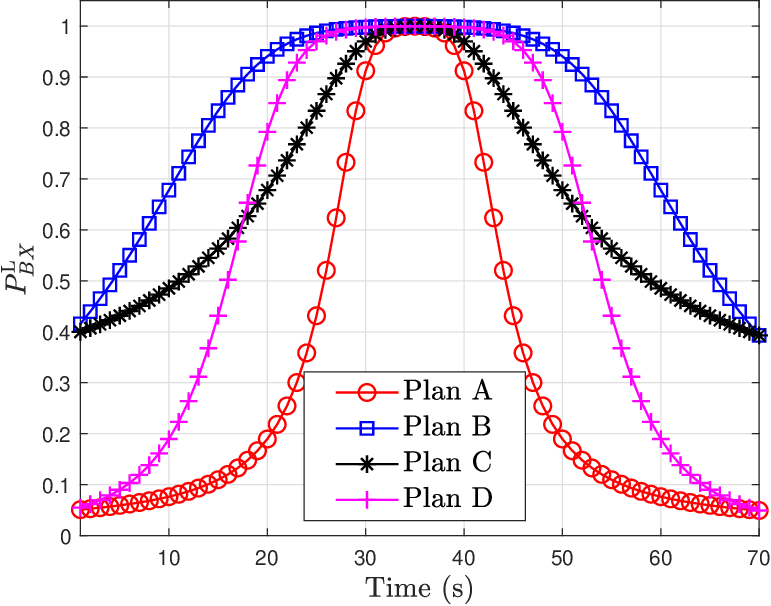}}
	\subfigure[The rate.]{
		\label{fig02c}
		\includegraphics[width = 0.23  \textwidth]{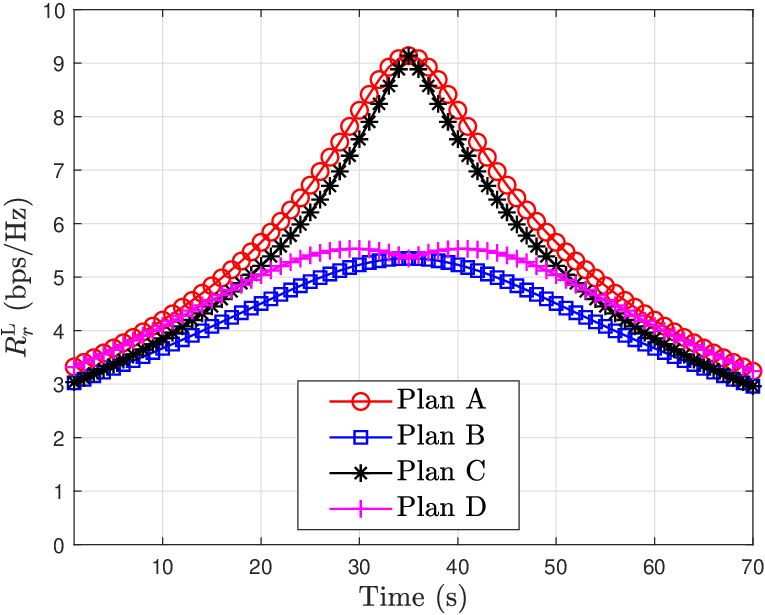}}
	\subfigure[The expected rate.]{
		\label{fig02d}
		\includegraphics[width = 0.23  \textwidth]{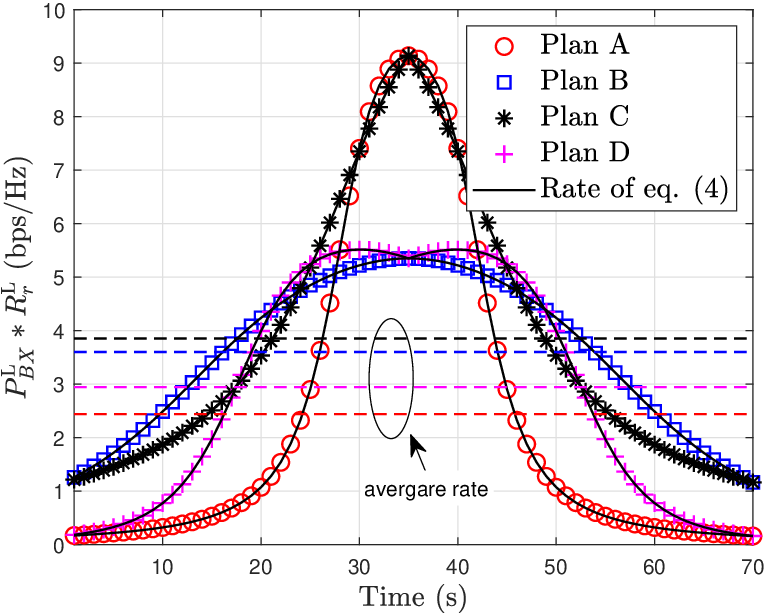}}
	\caption{{The comparison among different schemes.}}
	\label{fig02}
\end{figure*}

{As can be seen from (\ref{hBk}) and (\ref{expectedR1}), ${z_B}\left[ n \right]$ increases results in $d_{BX}\left[ n \right]$ increases, the path loss increases and the rate decreases. However, according to (\ref{eq1athetaBU}) and (\ref{eq1bthetaBD}), an increase of ${z_B}\left[ n \right]$ also leads to an increase of $ {\theta _{BX}}$, thus increasing $P_{BX}^{\mathrm{L}}\left[ n \right] $.
Thus, there is a tradeoff between the LoS probability and the rate, and there is an optimal altitude to maximize the expected rate given in (\ref{Rrlb}).
To illustrate this tradeoff, a simple scenario is illustrated in Fig. \ref{fig02}, in which the UAV flies from (-200,0) to (200,0) with different-vertical line trajectory shown in Fig. \ref{fig02a}, and the destination node is located at (0,0).
As can be seen from Fig. \ref{fig02b}, in the beginning, the LoS probability of the UAV located at 100 m is more significant than that at 30 m, and the probability of LoS increases when it approaches the user until it flies over the user. Plan A has the lowest LoS probability, and Plan B has the highest LoS probability.
It can be observed from Fig. \ref{fig02c} that Plan B has the lowest rate and Plan A has the highest rate at each slot due to the path loss.
Fig. \ref{fig02d} shows the expected rate with different schemes. We find that the results of (\ref{expectedR1}) and (\ref{Rrlb}) are almost equal, and the expected value of Plan c outperforms that of other schemes, which is consistent with the results given in Sec. \ref{Simulation}.}

The horizontal energy consumption of $B$ is expressed as \cite{ZengY2019TWC}
\begin{align}\label{speedxy}
		&P_{\mathrm {hor}}\left[ n \right] = \underbrace{P_0\left(1+\frac{3\left\|\mathbf{v}_{xy}\left[ n \right]\right\|^2}{U_{\mathrm {tip }}^2}\right)}_{\mathrm {blade profile }}+\underbrace{\frac{1}{2} d_0 \rho s A\left\|\mathbf{v}_{xy}\left[ n \right]\right\|^3}_{\mathrm {parasite }}  \nonumber\\
		&+\underbrace{P_1\left(\sqrt{1+\frac{\left\|\mathbf{v}_{xy}\left[ n \right]\right\|^4}{4 v_0^4}}-\frac{\left\|\mathbf{v}_{xy}\left[ n \right]\right\|^2}{2 v_0^2}\right)^{1/2}}_{\mathrm {induced }},
\end{align}
where
$\mathbf{v}_{xy}\left[ n \right]$ denotes the flying speed of $B$ in the hovering state,
$P_0$ and $P_1$ are two constants representing the inherent blade surface power and induced power, respectively, 
$U_{\mathrm {tip}}$ denotes the tip speed of the rotor blade,
$d_0$ and $\rho$ represent the body resistance ratio and air density, and the average rotor induced velocity is represented as $v_0$, $s$ and $A$ represent the rotor stiffness and rotor disc area, respectively.

The energy consumption of $B$ in the vertical direction is as expressed as \cite{MengA2022IoT, FilipponeABook}
\begin{align}\label{speedz}
	P_{\mathrm {ver}}\left[ n \right]=W v_{z}\left[ n \right], \quad \forall v_{z}\left[ n \right]>0,
\end{align}
where $W$ represents the weight of $B$.
UAVs consume no power during vertical descent \cite{FilipponeABook}. Therefore, when $v_{z}\left[ n \right] < 0$, we have $P_{\mathrm {ver}}\left[ n \right]=0$.

A binary variable $\alpha_r\left[ n \right]\in \{0,1\}$ is utilized  to represent whether $B$ sends information to $U_r$ in the $n$th time slot.
Specifically, $\alpha_r\left[ n \right]=1$ denotes that $U_r$  is scheduled, otherwise, $\alpha_r \left[ n \right]=0$.
It is assumed that $B$ can only communicate with one user in each time slot, then we have the following constraint
\begin{subequations}
	\begin{align}
		\sum\limits_{r=1}^R\alpha_r\left[ n \right] \leq 1, \forall n,   \label{diaodu1}\\
		\alpha_r\left[ n \right]\in \{0,1\}, \forall r,n.  \label{diaodu2}
	\end{align}
\end{subequations}
The  average rate of the considered system is expressed as
\begin{align}\label{AvgRate0}
	{\bar \Phi}  = \frac{1}{N}\sum\limits_{n = 1}^N {\Phi \left[ n \right]},
\end{align}
where
$\Phi \left[ n \right] = \sum\limits_{r = 1}^R {{\alpha _r}\left[ n \right]\bar R_r^{{\rm{lb}}}\left[ n \right]}$
signifies the sum rate of the considered system in each slot.

In underlay CRNs, the expected interference power constraint on $D$ is expressed as
\begin{align}\label{underlayCon}
	P_{BD}^{\mathrm{L}}\left[ n \right]\frac{P_B\left[ n \right] \gamma}{d_{BD}^{\alpha_L}\left[ n \right]} + P_{BD}^{\mathrm{N}}\left[ n \right]\frac{P_B\left[ n \right] \gamma \mu}{d_{BD}^{\alpha_N}\left[ n \right]} \leq \Gamma,
\end{align}
where $\Gamma$ denotes the peak interference power threshold value at $D$.

\section{Problem Formulation}
\label{Problem}

In this work,  the average rate of the system is optimized, which is related to user scheduling, the transmission power and 3D trajectory, the horizontal and vertical velocities of $B$.
Then the following optimization problem is formulated
\begin{subequations}
	\begin{align}
		\mathcal{P}_0: & \max _{\mathbf{A},\mathbf{P},\mathbf{\Theta}, \mathbf{Q}, \mathbf{H}} {\bar \Phi} \label{eq9a}\\
		\mathrm { s.t. }& 0 \leq P_B\left[ n \right] \leq P_B^{\max }, \forall n, \label{eq9b}\\
		& 0 \leq \frac{1}{N} \sum\limits_{n=1}^N P_B\left[ n \right] \leq {P_B^{\rm{ave}}},\label{eq9c} \\
		&\frac{1}{N} \sum\limits_{n=1}^N P_{\mathrm {hor }}\left[ n \right] \leq {P_{{\rm{hor}}}^{{\rm{ave}}}},\forall n, \label{eq9d}\\
		&\frac{1}{N} \sum\limits_{n=1}^N P_{\mathrm {ver }}\left[ n \right] \leq {P_{{\rm{ver}}}^{{\rm{ave}}}},\forall n, \label{eq9d2}\\
		&{\mathbf{q}}_B\left[ N \right] = {\mathbf{q}}_B\left[ 1 \right], \label{eq9e}\\
		& \left\|\mathbf{q}_B\left[ {n + 1} \right]-\mathbf{q}_B\left[ n \right]\right\| = \delta_t v_{xy}\left[ n \right], \forall n, \label{eq9f}\\
		&\left\|\mathbf{v}_{xy}\left[ n \right]\right\| \leq V_{\max }, \forall n, \label{eq9g} \\
		&\left\|\mathbf{v}_{xy}\left[ {n + 1} \right]-\mathbf{v}_{xy}\left[ n \right]\right\| \leq \delta_t  a_{\max }, \forall n,\label{eq9h}\\
		&{z_B}\left[ N \right] =  {z_B}\left[ 1 \right], \label{eq9i}\\
		& \left\|z_B\left[ {n + 1} \right]-z_B\left[ n \right]\right\| = \delta_t v_{z}\left[ n \right], \forall n, \label{eq9j}\\
		&\left\|{v}_{z}\left[ n \right]\right\| \leq \hat V_{\max }, \forall n, \label{eq9k} \\
		&H_{\min } \leq z_B\left[ n \right] \leq H_{\max }, \forall n, \label{eq9m}\\
		&  (\rm{\ref{eq1athetaBU}}),  (\rm{\ref{eq1bthetaBD}}), (\rm{\ref{diaodu1}}),   (\rm{\ref{diaodu2}}), (\rm{\ref{underlayCon}}).\nonumber
\end{align}
\end{subequations}
where
$\mathbf{A}=\left\{\alpha_r\left[ n \right], \forall r, n\right\}$,
$\mathbf{P}=\left\{P_B\left[ n \right], \forall n\right\}$,
$\mathbf{Q}=\{\mathbf{q}_B\left[ n \right], v_{xy}\left[ n \right],\forall n\}$,
$\mathbf{H}=\{z_B\left[ n \right], v_{z}\left[ n \right],\forall n\}$,
$\mathbf{\Theta}=\left\{\theta_{BU_r}\left[ {n} \right],\theta_{BD}\left[ {n} \right], \forall n,r\right\}$, 
$V^{max}$ and $\hat{V}^{max}$ represent the maximum horizontal and vertical velocity, respectively,
$a_ {\max}$ and $\hat a_{\max}$ represent the maximum horizontal and vertical acceleration, respectively,
$H_{\min}$ and $H_{\max}$ represent the minimum and maximum vertical altitude of $B$, respectively, 
and 
$P_B^{\max}$ and $P_B^{\rm{ave}} $ represent the maximum and average transmission power of $B$, respectively.
(\ref{eq9d}) and (\ref{eq9d2}) represent the propulsion energy constraint for $B$,
(\ref{eq9e}) and (\ref{eq9i}) are the constraints on the starting and final positions of $B$, that is, $B$ must fly back to the starting position at the end of each flight cycle,
(\ref{eq9f})-(\ref{eq9h}) and (\ref{eq9j})-(\ref{eq9m}) represent the velocity and acceleration constraints, respectively,
(\ref{eq9m}) denotes the flight altitude constraint of $B$,
({\ref{eq1athetaBU}}) and  ({\ref{eq1bthetaBD}}) are the constraints on elevation angle between $B$ and TN,
and
(\ref{diaodu1}) and (\ref{diaodu2}) signify the user scheduling constraints,
and
(\ref{underlayCon}) denotes the interference power constraint of $B$ on $D$,

Some factors make the optimization problem $\mathcal{P}_0$ difficult to solve. Firstly, the average rate expression in (\ref{Rrlb}) is related to the UAV's 3D trajectory and also to the probability of LoS/NLoS communication. 
Therefore, the objective function (\ref{eq9a}) in optimization problem $\mathcal{P}_0$ is highly coupled. The second constraint brought by user scheduling (\ref{diaodu2}) is binary integer constraint, which is non-convex. 
Also, consider that the elevation constraints (\ref{eq1athetaBU}) and (\ref{eq1bthetaBD}) caused by PLoS modeling are nonlinear constraints. 
Therefore, the optimization problem $\mathcal{P}_0$ is a nonlinear mixed integer non-convex optimization problem, which is usually troublesome to solve.

\section{Proposed Algorithm for Problem $\mathcal{P}_{0}$}
\label{ProposedAlgorithm1}

To solve $\mathcal {P}_0$, we utilize the BCD technology to decompose the original problem into multiple subproblems.
Specifically, for the given other variables, $\mathbf{A}$, $\mathbf{P}$, $\mathbf{H}$, and $\mathbf{Q}$ are optimized in each subproblem respectively.
In addition, the SCA technology is utilized to transform the non-convex constraints into convex constraints.

\subsection{Subproblem 1: Optimizing User Scheduling Variable }

In this subsection, the user scheduling variable $\mathbf{A}$ is optimized with given \{$\mathbf {P} $,$\mathbf{H} $,$\mathbf{Q}$,$\mathbf{\Theta}$\}.
The binary variable constraint $\alpha_r\left[ n \right] \in\{0,1\}$ is relaxed as a linear variable $0 \leq \alpha_r\left[ n \right] \leq 1$ and $\mathcal{P}_{0}$ is reformulated as
\begin{subequations}
	\begin{align}
		\mathcal{P}_{1.1}: & \max _{\mathbf{A},\eta} \eta \\
		\mathrm { s.t. } & {\bar \Phi} \geq \eta,  \label{p11b}\\
		& 0\leq\alpha_r\left[ n \right]\leq1, \forall n, r, \label{p11c}\\
		& (\rm{\ref{diaodu1}}).  \nonumber
	\end{align}
\end{subequations}
where $\eta$ is a slack variable. $\mathcal{P}_{1.1}$ is a strictly linear optimization problem, which can be effectively solved by using the CVX toolbox in \cite{tuyouhua}. 

\subsection{Subproblem 2: Optimizing Transmit Power of $B$ }

For given \{$\mathbf{A}$,$\mathbf{Q}$,$\mathbf{H}$,$\mathbf{\Theta}$\}, we have an optimization problem as follows
\begin{subequations}
	\begin{align}
		\mathcal{P}_{2.1}: & \max _{\mathbf{P},\eta} \eta \\
		\mathrm { s.t. } & {\bar \Phi} \geq \eta, \label{2.1}\\
		&   (\rm{\ref{underlayCon}}), (\rm{\ref{eq9b}}),  (\rm{\ref{eq9c}}).\nonumber
\end{align}\end{subequations}
The objective function in $\mathcal{P}_{2.1}$ is linear, the left-hand side (LHS) of  (\ref{2.1}) is a concave function in the form of $\log_2 (1+x)$, which is a convex constraint, and 
(\ref{underlayCon}), 
(\ref{eq9b}), and (\ref{eq9c}) are a linear constraints, so $\mathcal {P}_ {2.1} $ is a convex optimization problem and can be solved through the CVX toolbox.

\subsection{Subproblem 3: Optimizing Horizontal Trajectory and Velocity of $B$ }
In this subsection,  the horizontal trajectory and velocity of $B$ is optimized for provided \{$\mathbf{A}$,$\mathbf{P}$,$\mathbf{H}$\}. The original optimization problem is rewritten as
\begin{subequations}
	\begin{align}
		\mathcal{P}_{3.1}: & \max _{\mathbf{Q},\mathbf{\Theta},\eta} \eta \\
		\mathrm { s.t. } & {\bar \Phi} \geq \eta, \label{p31b}\\
		&   (\rm{\ref{underlayCon}}), (\rm{\ref{eq9d}}), (\rm{\ref{eq9e}})- (\rm{\ref{eq9h}}),   \nonumber \\
		&  (\rm{\ref{eq1athetaBU}}),  (\rm{\ref{eq1bthetaBD}}). \nonumber
		\end{align}
	\end{subequations}
$\mathcal {P}_ {3.1}$ is non-convex because (\ref{eq9d}) and  (\ref{underlayCon}) are non-convex constraints,
(\ref{eq1athetaBU}) and (\ref{eq1bthetaBD}) are nonlinear constraints,
and
$\bar{R}_{r}^{\mathrm{lb}}\left[ n \right]$ in constraint (\ref{p31b}) not only depends on $\mathbf{Q}$, but also on $P_{BU_r}^{\mathrm{L}}\left[ n \right]$.

\begin{figure*}[ht]
\begin{subequations}
	\begin{align}
		\frac{1}{N}\sum\limits_{n = 1}^N {\left[ {{P_0}\left( {1 + \frac{{3{{\left\| {{{\bf{v}}_{xy}}\left[ n \right]} \right\|}^2}}}{{U_{{\rm{tip}}}^2}}} \right) + \frac{1}{2}{d_0}\rho sA{{\left\| {{{\bf{v}}_{xy}}\left[ n \right]} \right\|}^3} + {P_1}{\lambda _B}\left[ n \right]} \right]}  \le P_{{\rm{hor}}}^{{\rm{ave}}} 	\label{shuipingslack} \\
		\frac{1}{\lambda_{{B}}\left[ n \right]^2} \leq \lambda_{{B}}\left[ n \right]^2+\frac{\left\|\mathbf{v}_{xy}\left[ n \right]\right\|^2}{v_0^2} 	\label{lambda}
	\end{align}
\end{subequations}
	\hrulefill
\end{figure*}
\setcounter{equation}{14}

By introducing a relaxation variable $\left\{\lambda_{{B}}\left[ n \right] \geq 0\right\}$, (\ref{eq9d}) is rewritten as (\ref{shuipingslack}) and (\ref{lambda}), shown at the top of this page,
It should be noted that the right-hand side (RHS) of (\ref{lambda}) is a convex function for both $\lambda_{{B}}\left[ n \right]$ and $\left\|\mathbf{v}_{xy}\left[ n \right]\right\|$, which means it is not a convex constraint.
By utilizing SCA and replacing the RHS with its convex lower bound, (\ref{lambda}) is rewritten as
\begin{align}\label{lambdasca}
		& \frac{1}{\lambda_{{B}}\left[ n \right]^2} \leq {\lambda}^{{\left( \kappa  \right)}}_{{B}}\left[ n \right]^2 + 2 {\lambda}^{{\left( \kappa  \right)}}_{{B}}\left[ n \right]\left(\lambda_{{B}}\left[ n \right]-{\lambda}^{{\left( \kappa  \right)}}_{{B}}\left[ n \right]\right)  \nonumber\\
		&+\frac{\left\|{\mathbf{v}}^{{\left( \kappa  \right)}}_{xy}\left[ n \right]\right\|^2}{v_0^2}+\frac{2}{v_0^2}\left({\bf{v}}^{{\left( \kappa  \right)}}_{xy}\left[ n \right]\right)^T\left({\bf{v}}_{xy}\left[ n \right]-{{\bf{v}}}^{{\left( \kappa  \right)}}_{xy}\left[ n \right]\right),
\end{align}
where
${\bf{v}}_{xy}^{{\left( \kappa  \right)}}\left[ n \right]$ represents the horizontal velocity obtained in the $k$th iteration.

Similarly, (\ref{underlayCon}) is reformulated as
 \begin{subequations}
 	\begin{align}
 		 \Gamma &\geq\frac{P_B\left[ n \right] \gamma}{p_{\mathrm{L}}\left[ n \right]t_{D,{\rm L}}\left[ n \right]} +\frac{P_B\left[ n \right] \gamma\mu}{p_{\mathrm{N}}\left[ n \right]t_{D, {\rm N}}\left[ n \right]} ,\label{eq15agama2}\\
 		p_\mathrm{L}\left[ n \right] &\leq 1+a e^{a b} e^{-b \theta_{BD}\left[ n \right]},\label{eq15bpl}\\
 		\theta_{BD}\left[ n \right] &\geq \frac{180}{\pi} \arctan \left(\frac{z_B\left[ n \right]}{\left\|\mathbf{q}_B\left[ n \right]-\mathbf{w}_D\right\|}\right),\label{eq15cthetabd1}\\
 		p_\mathrm{N}\left[ n \right] &\leq 1+\frac{1}a e^{-a b} e^{b \phi_{BD}\left[ n \right]},\label{eq15dpn}\\
 		\phi_{BD}\left[ n \right] &\leq \frac{180}{\pi} \arctan \left(\frac{z_B\left[ n \right]}{\left\|\mathbf{q}_B\left[ n \right]-\mathbf{w}_D\right\|}\right),\label{eq15ephibd} \\
 		t_{D,h}\left[ n \right] &\leq\left(\left\|\mathbf{q}_B\left[ n \right]-\mathbf{w}_{D}\right\|^2+z_B^2\left[ n \right]\right)^{\alpha_h / 2},\label{eq15ftdh}
 	\end{align}
 \end{subequations}
where
$h\in\{\mathrm{L},\mathrm{N}\}$,
$\mathbf{p}_h=\left\{p_h\left[ n \right],\forall n\right\}$,
and
$\mathbf{t}_{D, h}=\left\{t_{D, h}\left[ n \right], \forall n,r\right\}$ are relaxation variables,
(\ref{eq15cthetabd1}) and (\ref{eq15ephibd}) relax (\ref{eq1bthetaBD}) into inequality constraints.
It should be noted that the RHS of (\ref{eq15bpl}), (\ref{eq15dpn}), (\ref{eq15ephibd}) and  (\ref{eq15ftdh})  are convex functions, which make them non-convex constraints.
Then, the SCA technology is utilized to convert the terms
$e^{-b\theta_{BD}\left[ n \right]}$,
$e^{b\phi_{BD}\left[ n \right]}$,
$\arctan(\frac{z_B\left[ n \right]}{\|\mathbf{q}_B\left[ n \right]-\mathbf{w}_D\|})$
and
$(\|\mathbf{q}_B\left[ n \right] - \mathbf{w}_D\|^2+z_B^2\left[ n \right])^{\alpha_h/2}$
to linear functions.
By replacing them with their respective convex lower bounds, we obtain
\begin{subequations}
	\begin{align}
		e^{-b\theta_{BD}\left[ n \right]} &\geq F_{1}^{{\left( \kappa  \right)}}\left[ n \right], \label{f01} \\
		e^{b\phi_{BD}\left[ n \right]} &\geq F_{2}^{{\left( \kappa  \right)}}\left[ n \right],\label{f02} \\
		\left(\left\|\mathbf{q}_B\left[ n \right]-\mathbf{w}_D\right\|^2+z_B^2\left[ n \right]\right)^{\alpha_h/ 2} &\geq F_{3,h}^{{\left( \kappa  \right)}}\left[ n \right], \label{f03}\\
		\arctan \left(\frac{z_B\left[ n \right]}{\left\|\mathbf{q}_B\left[ n \right]-\mathbf{w}_D\right\|}\right) &\geq F_{5}^{{\left( \kappa  \right)}}\left[ n \right],\label{f05}
	\end{align}
\end{subequations}
where
$F_1^{\left( \kappa  \right)}\left[ n \right] = {e^{ - b\theta _{BD}^{\left( \kappa  \right)}\left[ n \right]}} - b{e^{ - b\theta _{BD}^{\left( \kappa  \right)}\left[ n \right]}}\left( {{\theta _{BD}}\left[ n \right] - \theta _{BD}^{\left( \kappa  \right)}\left[ n \right]} \right)$,
$F_2^{\left( \kappa  \right)}\left[ n \right] = {e^{b\phi _{BD}^{\left( \kappa  \right)}\left[ n \right]}} + b{e^{ - b\phi _{BD}^{\left( \kappa  \right)}\left[ n \right]}}\left( {{\phi _{BD}}\left[ n \right] - \phi _{BD}^{\left( \kappa  \right)}\left[ n \right]} \right)$,
$F_{3,h}^{\left( \kappa  \right)}\left[ n \right] = {\left( {d_{{B_D}}^{\left( \kappa  \right)}\left[ n \right]} \right)^{{\alpha _h}}} + {\alpha _h}{\left( {d_{{B_D}}^{\left( \kappa  \right)}\left[ n \right]} \right)^{{\alpha _h} - 2}} + {\left( {{\mathbf{q}}_B^{\left( \kappa  \right)}\left[ n \right] - {{\mathbf{w}}_D}} \right)^T}\left( {{{\mathbf{q}}_B}\left[ n \right] - {\mathbf{q}}_B^{\left( \kappa  \right)}\left[ n \right]} \right)$,
and
$F_5^{\left( \kappa  \right)}\left[ n \right] = \arctan \left( {\frac{{{z_B}\left[ n \right]}}{{\left\| {{\mathbf{q}}_B^{\left( \kappa  \right)}\left[ n \right] - {{\mathbf{w}}_D}} \right\|}}} \right) - \frac{{{z_B}\left[ n \right]\left( {\left\| {{{\mathbf{q}}_B}\left[ n \right] - {{\mathbf{w}}_D}} \right\| - \left\| {{\mathbf{q}}_B^{{\left( \kappa  \right)}}\left[ n \right] - {{\mathbf{w}}_D}} \right\|} \right)}}{{{{\left\| {{\mathbf{q}}_B^{\left( \kappa  \right)}\left[ n \right] - {{\mathbf{w}}_D}} \right\|}^2} + z_B^2\left[ n \right]}}$,
${\theta _{BD}^{\left( \kappa  \right)}\left[ n \right]}$,
${\phi _{BD}^{\left( \kappa  \right)}\left[ n \right]}$,
${d_{{B_D}}^{\left( \kappa  \right)}\left[ n \right]}$,
and
$\mathbf{q}_B^{{\left( \kappa  \right)}}\left[ n \right]$
denote ${\theta _{BD}\left[ n \right]}$,
${\phi _{BD}\left[ n \right]}$,
${d_{{B_D}}\left[ n \right]}$,
and
$\mathbf{q}_B\left[ n \right]$  in the $k$th iteration, respectively.
Therefore, (\ref{eq15bpl}), (\ref{eq15dpn}), (\ref{eq15ephibd}), and (\ref{eq15ftdh}) are reformulated as
\begin{subequations}
	\begin{align}
		p_\mathrm{L}\left[ n \right] &\leq 1+a e^{ab}F_{1}^{{\left( \kappa  \right)}}\left[ n \right], \label{17aF1}\\
		p_\mathrm{N}\left[ n \right] &\leq 1+\frac{1}{a} e^{-ab}F_{2}^{{\left( \kappa  \right)}}\left[ n \right],\label{17bF2} \\
		\phi_{BD}\left[ n \right] &\leq \frac{180}{\pi}F_{5}^{{\left( \kappa  \right)}}\left[ n \right],\label{17cF5} \\
		t_{D, h}\left[ n \right] &\leq F_{3,h}^{{\left( \kappa  \right)}}\left[ n \right].\label{17dF3}
	\end{align}
\end{subequations}

By introducing slack variables $\mathbf{x}_L=\left\{x_L\left[ n \right],\forall n\right\}$ and $\mathbf{t}_{r, L}=\left\{t_{r, L}\left[ n \right], \forall n,r\right\}$,
(\ref{p31b}) is reformulated as
\begin{subequations}
	\begin{align}
		& {\bar \Phi_1} \geq \eta,  \label{p32a}\\
		&x_L\left[ n \right] \geq 1+a e^{a b} e^{-b \theta_{BU_r}\left[ n \right]},\label{p32b} \\
		&\theta_{BU_r}\left[ n \right] \leq \frac{180}{\pi} \arctan \left(\frac{z_B\left[ n \right]}{\left\|\mathbf{q}_B\left[ n \right]-\mathbf{w}_{U_r}\right\|}\right), \label{p32c} \\
		&t_{r,L}\left[ n \right] \geq\left(\left\|\mathbf{q}_B\left[ n \right]-\mathbf{w}_{U_r}\right\|^2+z_B^2\left[ n \right]\right)^{\alpha_L / 2}, \label{p32d}
	\end{align}
\end{subequations}
where
${\bar \Phi_1} = \frac{1}{N} \sum\limits_{n=1}^N \sum\limits_{r=1}^R\alpha_r\left[ n \right] \bar{R}_{r,1}\left[ n \right] $,
$\bar{R}_{r,1}\left[ n \right]  =\frac{1}{x_L\left[ n \right]} \log _2\left(1+\frac{P_B\left[ n \right] \gamma}{t_{r, L}\left[ n \right]}\right)$.
It should be noted that (\ref{p32c}) relaxes (\ref{eq1athetaBU}) into an inequality constraint.
With the same method as (\ref{eq15bpl}), we have
\begin{align}\label{F4}
	\theta_{BU_r}\left[ n \right] \leq \frac{180}{\pi}F_4^{{\left( \kappa  \right)}}\left[ n \right], 
\end{align}
where
$F_4^{\left( \kappa  \right)}\left[ n \right] = \arctan \left( {\frac{{{z_B}\left[ n \right]}}{{\left\| {{\mathbf{q}}_B^{\left( \kappa  \right)}\left[ n \right] - {{\mathbf{w}}_{{U_r}}}} \right\|}}} \right) - \frac{{{z_B}\left[ n \right]\left( {\left\| {{{\mathbf{q}}_B}\left[ n \right] - {{\mathbf{w}}_{{U_r}}}} \right\| - \left\| {{\mathbf{q}}_B^{\left( \kappa  \right)}\left[ n \right] - {{\mathbf{w}}_{{U_r}}}} \right\|} \right)}}{{{{\left\| {{\mathbf{q}}_B^{\left( \kappa  \right)}\left[ n \right] - {{\mathbf{w}}_{{U_r}}}} \right\|}^2} + z_B^2\left[ n \right]}}$.

To address the non-convexity in (\ref{p32a}), \textit{Lemma 1} is introduced.

\emph{Lemma 1}: Given any $A\geq0$, the function $f\left(x,y\right)=$ $\frac{1}{x} \log _2\left(1+\frac{A}{y}\right)$ is jointly convex with respect to $x > 0$ and $y>0$.

\emph{Proof}: See the Appendix \ref{lemma1}.

Based on $\emph{Lemma 1}$, the first-order Taylor expansion of $\bar{R}_{r,1}\left[ n \right]$ is obtained as
\begin{align}\label{Rr1lb}
		\bar{R}_{r,1}^{\mathrm{lb}}\left[ n \right] & = A_{r}^{{\left( \kappa  \right)}}\left[ n \right]+B_{r,L}^{{\left( \kappa  \right)}}\left[ n \right]\left(t_{r, L}\left[ n \right]-t_{r, L}^{{\left( \kappa  \right)}}\left[ n \right]\right) \nonumber\\
		& \,\,\,\,\,\,\, +C_{r, L}^{{\left( \kappa  \right)}}\left[ n \right]\left(x_L\left[ n \right]-x_L^{{\left( \kappa  \right)}}\left[ n \right]\right),
\end{align}
where
$A_r^{{\left( \kappa  \right)}}\left[ {n} \right] = \frac{{{{\log }_2}\left( {1 + \frac{{{P_B}\left[ {n} \right]\gamma }}{{t_{r,{\mathrm{L}}}^{{\left( \kappa  \right)}}\left[ {n} \right]}}} \right)}}{{x_{\mathrm{L}}^{{\left( \kappa  \right)}}\left[ {n} \right]}}$,
$B_{r,L}^{{\left( \kappa  \right)}}\left[ n \right]  =\frac{-P_B\left[ n \right] \gamma}{\ln (2) x_L^{{\left( \kappa  \right)}}\left[ n \right]\left(t_{r,L}^{{\left( \kappa  \right)}}\left[ n \right]\right)^2\left(1+\frac{P_B\left[ n \right] \gamma}{t_{r, L}^{{\left( \kappa  \right)}}\left[ n \right]}\right)}$,
$C_{r,L}^{{\left( \kappa  \right)}}\left[ n \right]  =-\frac{\log _2\left(1+\frac{P_B\left[ n \right] \gamma}{t_{r, L}^{{\left( \kappa  \right)}}\left[ n \right]}\right)}{\left(x_L^{{\left( \kappa  \right)}}\left[ n \right]\right)^2}$,
and
$x_L^{{\left( \kappa  \right)}}\left[ n \right]$, $t_{r,L}^{{\left( \kappa  \right)}}\left[ n \right]$ and $t_{D,h}^{{\left( \kappa  \right)}}\left[ n \right]$ are $x_{\mathrm{L}}\left[ n \right]$, $t_{r,L}\left[ n \right]$ and $t_{D,h}\left[ n \right]$ at the $k$th iteration, respectively.

Finally, $\mathcal{P}_{3.1}$ is rewritten as
\begin{subequations}
	\begin{align}
		\mathcal{P}_{3.2} & \max _{\mathbf{Q}, \boldsymbol{\Theta}, \eta, \lambda_{B}, {x}_{\mathrm{L}}, \mathbf{p}_h, \mathbf{t}_{r,L}, \mathbf{t}_{D, h}} \eta \\
		\mathrm { s.t. } & {\bar \Phi_2} \geq \eta, \\
		& (\rm{\ref{eq9e}})- (\rm{\ref{eq9h}}), (\rm{\ref{shuipingslack}}), (\rm{\ref{lambdasca}}), (\rm{\ref{eq15agama2}}),  (\rm{\ref{eq15cthetabd1}}),  \nonumber\\
		& (\rm{\ref{17aF1}})-(\rm{\ref{17dF3}}), (\rm{\ref{p32b}}), (\rm{\ref{p32d}}), (\rm{\ref{F4}}), \nonumber
	\end{align}
\end{subequations}
where
${\bar \Phi_2} = \frac{1}{N} \sum\limits_{n=1}^N \sum\limits_{r=1}^R\alpha_r\left[ n \right] \bar{R}_{r, 1}^{\mathrm{lb}}\left[ n \right]$.
$\mathcal{P}_{3.2}$ is a convex problem that can be solved using existing optimization tools such as CVX.

\subsection{Subproblem 4: Optimizing Horizontal Trajectory and Velocity of $B$ }

In this subsection, for given \{$\mathbf{A}$,$\mathbf{P}$,$\mathbf{Q}$\},  the vertical trajectory $\mathbf{H}$ of $B$ is optimized. The optimization problem is expressed as
\begin{subequations}
	\begin{align}
		\mathcal{P}_{4.1} & : \max _{\mathbf{H},\mathbf{\Theta}, \eta} \eta \\
		\mathrm { s.t. } & {\bar \Phi} \geq \eta, \label{eqp41b}\\
		&  (\rm{\ref{underlayCon}}),  (\rm{\ref{eq9d2}}), (\rm{\ref{eq9i}})-(\rm{\ref{eq9m}}), (\rm{\ref{eq1athetaBU}}), (\rm{\ref{eq1bthetaBD}}). \nonumber
	\end{align}
\end{subequations}

It should be noted that $\mathcal{P}_{4.1}$ has the same problems as $\mathcal{P}_{3.1}$.
Specifically, (\ref{eqp41b}) not only depends on $\mathbf{H}$, but also on $P_{BU_r}^{\mathrm{L}}\left[ n \right]$,
({\ref{underlayCon}}) is non-convex constraint,
and
({\ref{eq1athetaBU}}) and ({\ref{eq1bthetaBD}}) are nonlinear constraints.

With the same method as (\ref{p31b}), (\ref{eqp41b}) is reformulated as (\ref{p32a})-(\ref{p32d}) and ({\ref{eq1athetaBU}}) and ({\ref{eq1bthetaBD}}) are reformulated as  ({\ref{eq15cthetabd1}}), ({\ref{eq15ephibd}}), and ({\ref{p32c}}).
With the same method in \textit{Subproblem 3}, (\ref{underlayCon}) in this subsection is reformulated as ({\ref{eq15agama2}})-({\ref{eq15ftdh}}) wherein ({\ref{eq15bpl}}) and ({\ref{eq15dpn}}) are  reformulated as ({\ref{17aF1}}) and ({\ref{17bF2}}), respectively.

The SCA is utilized to deal with  ({\ref{eq15cthetabd1}})  and ({\ref{eq15ftdh}}) with {respect} to
$(\|\mathbf{q}_B\left[ n \right]-\mathbf{w}_D\|^2+z_B^2\left[ n \right])^{\alpha_h/2}$
and
$\arctan(\frac{z_B\left[ n \right]}{\|\mathbf{q}_B\left[ n \right]-\mathbf{w}_D\|})$, respectively.
Similarly, we obtain
\begin{subequations}
	\begin{align}
		{\left( {{{\left\| {{{\mathbf{q}}_B}\left[ {n} \right] - {{\mathbf{w}}_D}} \right\|}^2} + z_B^2\left[ {n} \right]} \right)^{{\alpha _h}/2}} \geqslant F_{6,h}^{(\kappa )}\left[ {n} \right], \label{F06}\\
		\arctan \left( {\frac{{{z_B}\left[ {n} \right]}}{{\left\| {{{\mathbf{q}}_B}\left[ {n} \right] - {{\mathbf{w}}_D}} \right\|}}} \right) \geqslant F_7^{(\kappa )}\left[ {n} \right], \label{F08}
	\end{align}
\end{subequations}
where
$  F_{6,h}^{(\kappa )}\left[ {n} \right] = {\left( {d_{BD}^{(\kappa )}\left[ {n} \right]} \right)^{{\alpha _h}}} + {\alpha _h}{\left( {d_{BD}^{(\kappa )}\left[ {n} \right]} \right)^{{\alpha _h} - 2}}{\left( {z_B^{(\kappa )}\left[ {n} \right]} \right)^T}\left( {{z_B}\left[ {n} \right] - z_B^{(\kappa )}\left[ {n} \right]} \right) $,
$F_7^{(\kappa )}\left[ {n} \right] = $ $\arctan \left( {\frac{{z_B^{(\kappa )}\left[ {n} \right]}}{{\left\| {{{\mathbf{q}}_B}\left[ {n} \right] - {{\mathbf{w}}_D}} \right\|}}} \right) + \frac{{\left\| {{{\mathbf{q}}_B}\left[ {n} \right] - {{\mathbf{w}}_D}} \right\|\left( {{z_B}\left[ {n} \right] - z_B^{(\kappa )}\left[ {n} \right]} \right)}}{{{{\left\| {{{\mathbf{q}}_B}\left[ {n} \right] - {{\mathbf{w}}_D}} \right\|}^2} + z_B^{(\kappa )}{{\left[ {n} \right]}^2}}}$,
and
$z_B^{(\kappa )}\left[ n \right] $ denotes ${z_B}\left[ n \right]$
in the $k$th iteration.
Finally, ({\ref{eq15cthetabd1}})  and ({\ref{eq15ftdh}})  are reformulated as
\begin{subequations}
	\begin{align}
		t_{D, h}\left[ n \right] &\leq F_{6, h}^{{\left( \kappa  \right)}}\left[ n \right], \label{F6}\\
		\theta_{BD}\left[ n \right] &\geq \frac{180}{\pi} F_7^{{\left( \kappa  \right)}}\left[ n \right]. \label{F7}
	\end{align}
\end{subequations}

Then $\mathcal{P}_{4.1}$ is rewritten as
\begin{subequations}
	\begin{align}
		\mathcal{P}_{4.2} & \max _{\mathbf{H}, \mathbf{\Theta},\eta, \mathbf{x}_{L}, \mathbf{p}_h, \mathbf{t}_{r,L}, \mathbf{t}_{D,h}} \eta \\
		\mathrm { s.t. } &{\bar \Phi_2} \geq \eta , \\
		& (\rm{\ref{eq9d2}}), (\rm{\ref{eq9i}})-(\rm{\ref{eq9m}}), (\rm{\ref{eq15agama2}}), (\rm{\ref{eq15ephibd}}),  (\rm{\ref{17aF1}}), \nonumber\\
		& (\rm{\ref{17bF2}}), (\rm{\ref{p32b}})-(\rm{\ref{p32d}}), (\rm{\ref{F6}}),(\rm{\ref{F7}}). \nonumber
	\end{align}
\end{subequations}
$\mathcal {P}_{4.2}$ is a convex optimization problem that can be solved using existing optimization tools such as CVX.

\subsection{Convergence Analysis of Algorithm 1}

$\mathcal {P}_0$ is a non-convex optimization problem, which decouples the variables $\mathbf {A} $, $\mathbf{P} $, $\mathbf{\Theta}$, $\mathbf{H}$, and $\mathbf{Q}$ in the original problem.
Through BCD, solving $\mathcal {P}_0$ converts to solving $\mathcal {P}_ {1.1} $, $\mathcal {P}_ {2.1} $, $\mathcal {P}_ {3.1}$ and $\mathcal {P}_ {4.1}$ in turn.
By using SCA and introducing relaxation variables, $\mathcal {P}_ {3.1} $ and $\mathcal {P}_ {4.1} $ that have  non-convex constraints
are rewritten as $\mathcal {P}_ {3.2} $ and $\mathcal {P}_ {4.2}$, which are all convex optimization problems.
The value of the objective function for the original problem $\mathcal{P}_0$ is defined as $R\left(\mathbf{A}^{\kappa},\mathbf{P}^{\kappa},\mathbf{\Theta}^{\kappa},\mathbf{H}^{\kappa}, \mathbf{Q}^{\kappa}\right)$ at the $k$th iteration,
\textbf{Algorithm 1} summarizes the details of the overall iteration for solving problem $\mathcal{P}_0$, where $\varepsilon $ represents the exact convergence value.

The obtained suboptimal solution of the transformed subproblem is also the suboptimal solution of the original non-convex subproblem, and each subproblem is solved using SCA convex transformation iteration. Finally, all suboptimal solutions of the subproblems that satisfy the threshold $\varepsilon$ constitute the suboptimal solution of the original problem. Therefore, our algorithm is to alternately solve the subproblem $\mathcal{P}_ {1.1} $, $\mathcal{P}_ {2.1} $, $\mathcal{P}_ {3.2} $ and $\mathcal{P}_ {4.2} $ to obtain the suboptimal solution of the original problem until a solution that satisfies the threshold $\varepsilon $ is obtained.

It is worth noting that in the classic BCD, to ensure the convergence of the algorithm, it is necessary to accurately solve and update the subproblems of each variable block with optimality in each iteration. But when we solve $\mathcal{P}_ {3.1} $ and $\mathcal{P}_ {4.1}$ , we can only optimally solve their approximation problem $\mathcal{P}_ {3.2} $ and $\mathcal{P}_ {4.2}$.
Therefore, we cannot directly apply the convergence analysis of the classical BCD, and further proof of the convergence of \textbf{Algorithm 1} is needed, as shown below.

\begin{algorithm}[t]
	\caption{Iterative Procedure of Problem $\mathcal{P}_{0}$}
	\KwIn{Initialization of feasible points.}
\While{$R( \mathbf{A}^{\kappa}, \mathbf{P}^{\kappa}, \mathbf{\Theta}^{\kappa},  \mathbf{H}^{\kappa}, \mathbf{Q}^{\kappa}) - $\\
$R( \mathbf{A}^{\kappa-1}, \mathbf{P}^{\kappa-1},  \mathbf{\Theta}^{\kappa-1},  \mathbf{H}^{\kappa-1}, \mathbf{Q}^{\kappa-1}) \succ \varepsilon$}{
Solve $\left(\mathcal{P}_{1.1}\right)$ for given $\left\{\mathbf{P}^{\kappa},\mathbf{H}^{\kappa},\mathbf{Q}^{\kappa},\mathbf{\Theta}^{\kappa}\right\}$\\
 and obtain the solution $\mathbf{A}^{\kappa+1}$;\\
Solve $\left(\mathcal{P}_{2.1}\right)$ for given $\left\{\mathbf{A}^{\kappa+1},\mathbf{H}^{\kappa},\mathbf{Q}^{\kappa},\mathbf{\Theta}^{\kappa}\right\}$ \\
and obtain the solution $\mathbf{P}^{\kappa+1}$;\\
Solve $\left(\mathcal{P}_{3.2}\right)$ for given $\left\{\mathbf{A}^{\kappa+1},\mathbf{H}^{\kappa},\mathbf{P}^{\kappa+1}\right\}$ and
\\ obtain the solution $\mathbf{Q}^{\kappa+1}$;\\
Solve $\left(\mathcal{P}_{4.2}\right)$ for given $\left\{\mathbf{A}^{\kappa+1},\mathbf{Q}^{\kappa+1},\mathbf{P}^{\kappa+1}\right\}$ and \\ obtain the solution $\mathbf{H}^{\kappa+1}$;\\
$\kappa=\kappa+1$;\\
Compute the objective value\\
$R( \mathbf{A}^{\kappa}, \mathbf{P}^{\kappa}, \mathbf{\Theta}^{\kappa},  \mathbf{H}^{\kappa}, \mathbf{Q}^{\kappa})$.}
\KwOut{$R( \mathbf{A}^{\kappa}, \mathbf{P}^{\kappa}, \mathbf{\Theta}^{\kappa},  \mathbf{H}^{\kappa}, \mathbf{Q}^{\kappa})$ with
$\mathbf{A}^*=\mathbf{A}^{\kappa}$, \\
$\mathbf{P}^*=\mathbf{P}^{\kappa}$,$\mathbf{\Theta}^*=\mathbf{\Theta}^{\kappa}$, $\mathbf{H}^*=\mathbf{H}^{\kappa}$,$\mathbf{Q}^*=\mathbf{Q}^{\kappa}$.}
\end{algorithm}

Firstly, for step 1 of our proposed \textbf{Algorithm 1}, we obtain the optimal solution to user scheduling problem $\mathcal{P}_{1.1}$ under the condition of given other feasible variables $\mathbf{P}^\kappa$, $\mathbf{H}^\kappa$, $\mathbf{Q}^\kappa$ and $\mathbf{\Theta}^\kappa$, therefore we can obtain
\begin{align}\label{z1.1}
	R\left(\mathbf{A}^\kappa, \mathbf{P}^\kappa, \mathbf{\Theta}^\kappa, \mathbf{H}^\kappa, \mathbf{Q}^\kappa\right)\leq R\left(\mathbf{A}^{\kappa+1}, \mathbf{P}^{\kappa}, \mathbf{\Theta}^{\kappa}, \mathbf{H}^{\kappa}, \mathbf{Q}^{\kappa}\right). 
\end{align}

Similarly, step 2 in \textbf{Algorithm 1} is to solve the power optimization problem $\mathcal {P}_ {2.1} $, given other feasible variables $\mathbf{A}^{\kappa+1}$, $\mathbf{H}^\kappa$, $\mathbf{Q}^\kappa$ and $\mathbf{\Theta}^\kappa$, this problem is a convex optimization problem, so we can also obtain the optimal solution for $\mathcal {P}_ {2.1} $, so we have
\begin{align}\label{z2.1}
	R\left(\mathbf{A}^{\kappa+1}, \mathbf{P}^\kappa, \mathbf{\Theta}^\kappa, \mathbf{H}^\kappa, \mathbf{Q}^\kappa\right)\leq R\left(\mathbf{A}^{\kappa+1}, \mathbf{P}^{\kappa+1}, \mathbf{\Theta}^{\kappa}, \mathbf{H}^{\kappa}, \mathbf{Q}^{\kappa}\right). 
\end{align}

Then, in step 3 of \textbf{Algorithm 1}, the approximate problem $\mathcal{P}_{3.2}$ of Horizontal Trajectory Optimization Problem $\mathcal{P}_{3.2}$ is solved by fixing other feasible variables $\mathbf{A}^{\kappa+1}$, $\mathbf{P}^{\kappa+1}$ and $\mathbf{H}^\kappa$. We obtain
\begin{align}\label{z3.2}
		&R\left(\mathbf{A}^{\kappa+1}, \mathbf{P}^{\kappa+1}, \mathbf{\Theta}^{\kappa}, \mathbf{H}^{\kappa}, \mathbf{Q}^{\kappa}\right) \nonumber\\
		& \stackrel{(a)}{=} R^{lb}\left(\mathbf{A}^{\kappa+1}, \mathbf{P}^{\kappa+1}, \mathbf{\Theta}^{\kappa}, \mathbf{H}^{\kappa}, \mathbf{Q}^{\kappa}\right) \nonumber\\
		& \stackrel{(b)}{\leq} R^{lb}\left(\mathbf{A}^{\kappa+1}, \mathbf{P}^{\kappa+1}, \mathbf{\Theta}^{\kappa}, \mathbf{H}^{\kappa}, \mathbf{Q}^{\kappa+1}\right) \nonumber\\
		& \stackrel{(c)}{\leq} R\left(\mathbf{A}^{\kappa+1}, \mathbf{P}^{\kappa+1}, \mathbf{\Theta}^{\kappa}, \mathbf{H}^{\kappa}, \mathbf{Q}^{\kappa+1}\right). 
\end{align}

Among them, (a) holds because given $\mathbf{Q}^{\kappa}$, $\mathcal{P}_{3.1}$ and $\mathcal{P}_{3.2}$ has the same objective function value; (b) is bounded because given $\mathbf{A}^{\kappa+1}$, $\mathbf{H}^{\kappa}$ and $\mathbf{P}^{\kappa+1}$, $\mathbf{Q}^{\kappa+1} $ is maximizing $R^{\rm{lb}}$ obtained from $\mathcal{P}_{3.2}$; (c) is established because the target value of problem $\mathcal{P}_{3.2}$ is the lower bound of the target value of its original problem $\mathcal{P}_{3.2}$ at $\mathbf{Q}^{\kappa+1}$. The inequality in (\ref{z3.2}) indicates that although the UAV trajectory was obtained by solving approximate optimization problem, the objective value of problem $\mathcal{P}_{3.2}$ still does not decrease after each iteration. Finally, for step 4 in \textbf{Algorithm 1}, it is also achieved by fixing other feasible variables $\mathbf{A}^{\kappa+1}$, $\mathbf{Q}^{\kappa+1}$ and $\mathbf{P}^{\kappa+1}$, similar to step 3, we obtain
\begin{align}\label{z4.2}
		&R\left(\mathbf{A}^{\kappa+1}, \mathbf{P}^{\kappa+1}, \mathbf{\Theta}^\kappa, \mathbf{H}^\kappa, \mathbf{Q}^{\kappa+1}\right) \nonumber\\
		& \stackrel{(a)}{=} R^{l b}\left(\mathbf{A}^{\kappa+1}, \mathbf{P}^{\kappa+1}, \mathbf{\Theta}^\kappa, \mathbf{H}^\kappa, \mathbf{Q}^{\kappa+1}\right) \nonumber\\
		& \stackrel{(b)}{\leq} R^{l b}\left(\mathbf{A}^{\kappa+1}, \mathbf{P}^{\kappa+1}, \boldsymbol{\Theta}^{\kappa+1}, \mathbf{H}^{\kappa+1}, \mathbf{Q}^{\kappa+1}\right) \nonumber\\
		& \stackrel{(c)}{\leq} R\left(\mathbf{A}^{\kappa+1}, \mathbf{P}^{\kappa+1}, \mathbf{\Theta}^{\kappa+1}, \mathbf{H}^{\kappa+1}, \mathbf{Q}^{\kappa+1}\right).
\end{align}

This is similar to the representation in (\ref{z3.2}), and from (\ref{z1.1}) to (\ref{z4.2}), we obtain
\begin{align}\label{z4.3}
	&R\left(\mathbf{A}^{\kappa}, \mathbf{P}^{\kappa}, \mathbf{\Theta}^\kappa, \mathbf{H}^\kappa, \mathbf{Q}^{\kappa}\right) \nonumber\\
	&\leq R\left(\mathbf{A}^{\kappa+1}, \mathbf{P}^{\kappa+1}, \mathbf{\Theta}^{\kappa+1}, \mathbf{H}^{\kappa+1}, \mathbf{Q}^{\kappa+1}\right).
\end{align}
The above analysis indicates that the target value of $\mathcal{P}_{0}$ does not decrease after each iteration of Algorithm 1. 
Due to the objective value of $\mathcal{P}_{0}$ is a finite upper bound, therefore the proposed \textbf{Algorithm 1} ensures convergence. 
The simulation results in the next section indicate that the proposed BCD-based method converges rapidly for the setting we are considering. In addition, since only convex optimization problems need to be solved in each iteration of \textbf{Algorithm 1}, which have polynomial complexity, \textbf{Algorithm 1} can actually converge quickly for wireless networks with a moderate number of users.

In summary, by alternately solving the four subproblems $\mathcal {P}_{1.1} $, $\mathcal {P}_{2.1} $, $\mathcal {P}_ {3.2} $ and $\mathcal{P}_ {4.2} $, until the algorithm converges to the predetermined accuracy $\varepsilon>0 $, we can obtain the suboptimal solution to problem $\mathcal{P}_{0}$. In addition, the computational complexity of this algorithm is $O\left(N^{3.5} \log (1 / \varepsilon)\right)$.

\begin{figure*}[t]
	\centering
	\subfigure[The average rate with different schemes.]{
		\label{fig03a}
		\includegraphics[width = 0.31  \textwidth]{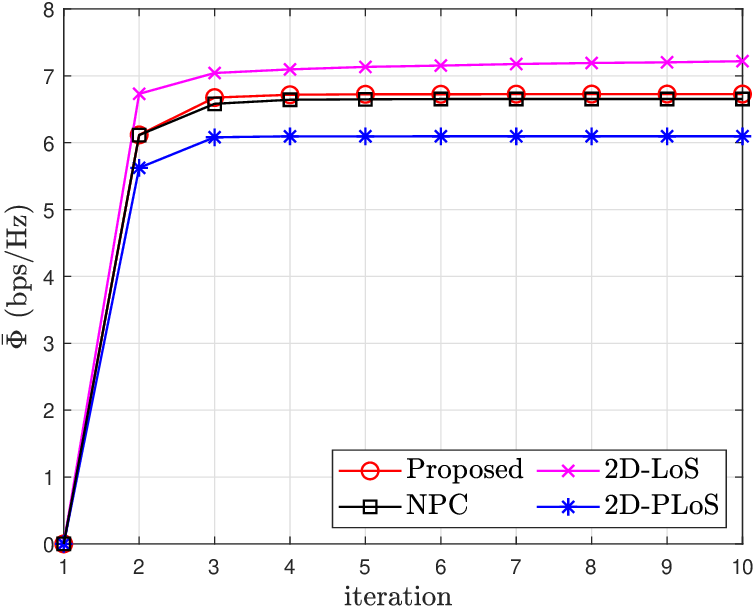}}
	\subfigure[The average rate with different power limitation.]{
		\label{fig03b}
		\includegraphics[width = 0.31  \textwidth]{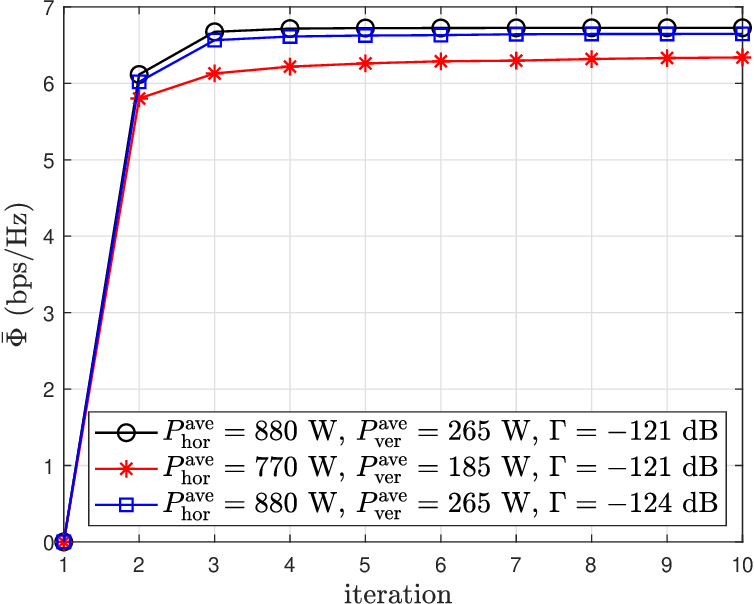}}
	\subfigure[The user scheduling.]{
		\label{fig03c}
		\includegraphics[width = 0.31  \textwidth]{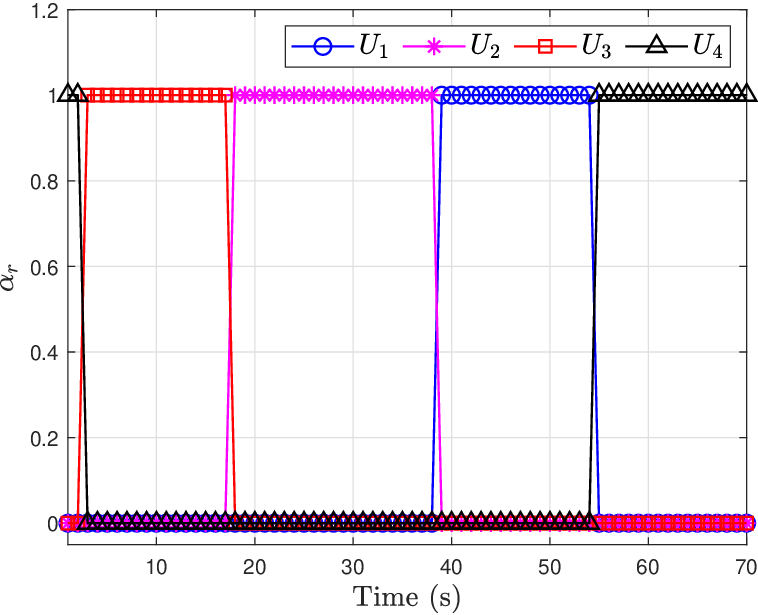}}
	\caption{{The average rate and user scheduling.}}
	\label{fig03}
\end{figure*}
\begin{figure*}[t]
	\centering
	\subfigure[3D trajectories of $B$ with different schemes.]{
		\label{fig04a}
		\includegraphics[width = 0.4 \textwidth]{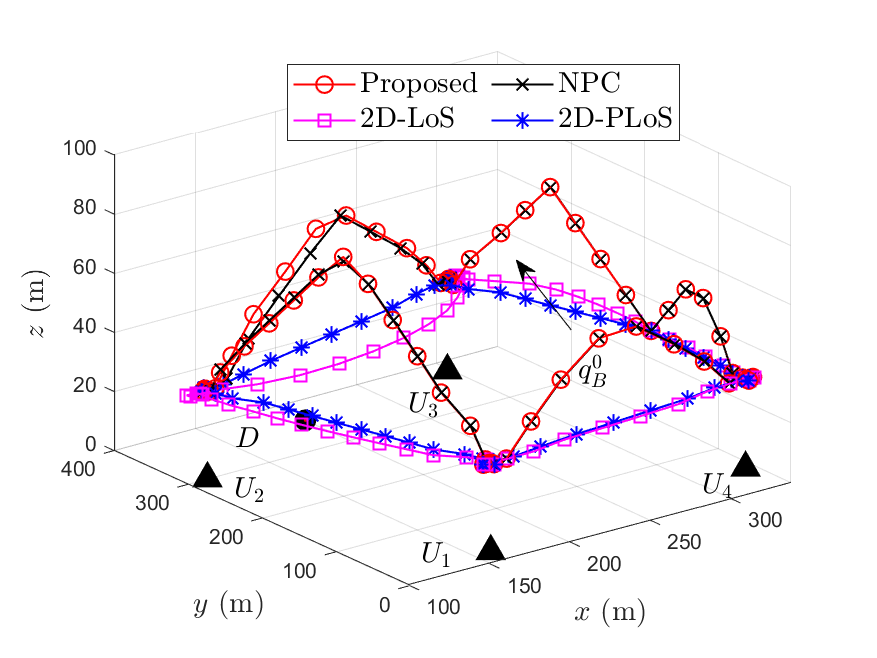}}
	\subfigure[3D trajectories of $B$ under different scenarios.]{
		\label{fig04b}
		\includegraphics[width = 0.4 \textwidth]{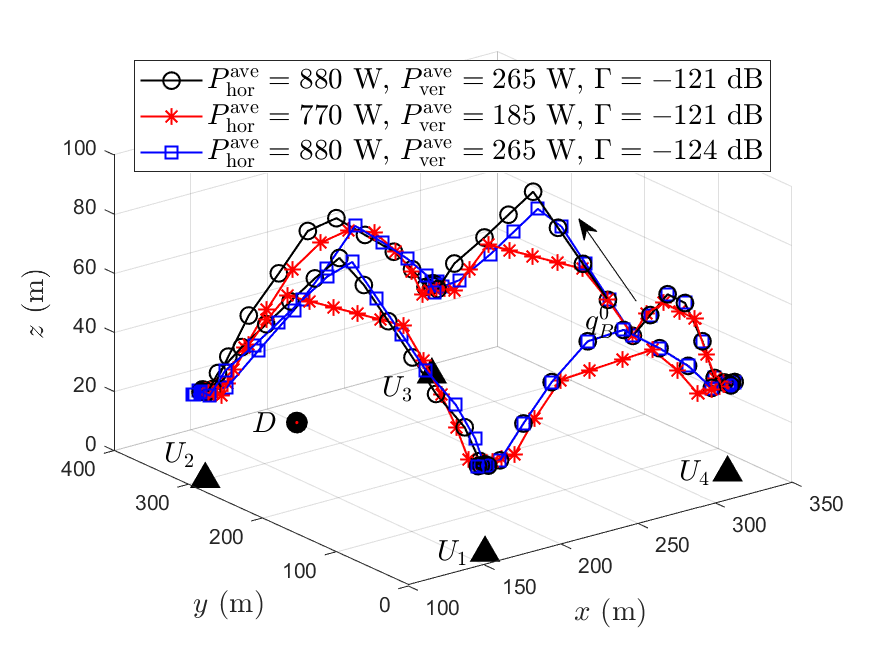}}
	\caption{3D trajectories of $B$ under different schemes and scenarios.}
	\label{fig04}
\end{figure*}
\begin{figure*}[t]
	\centering
	\subfigure[The horizontal trajectory of $B$.]{
		\label{fig05a}
		\includegraphics[width = 0.31  \textwidth]{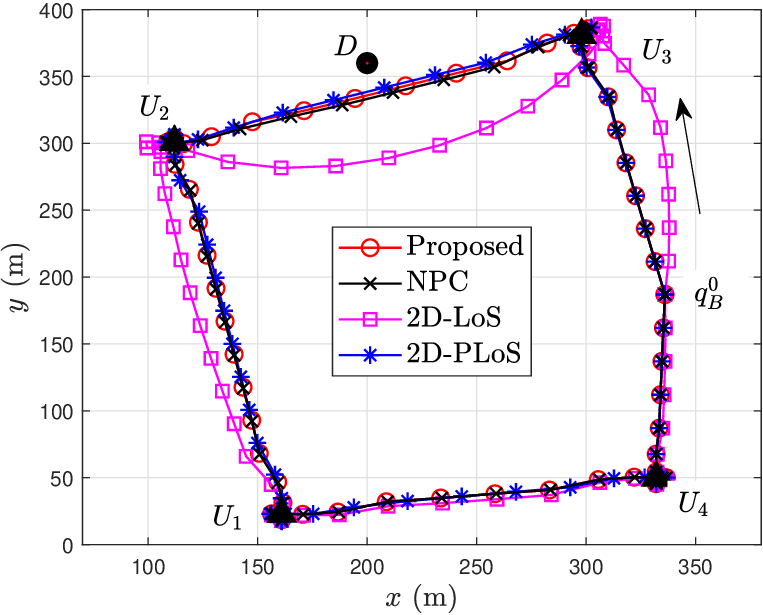}}
	\subfigure[The vertical attitude of $B$.]{
		\label{fig05b}
		\includegraphics[width = 0.31  \textwidth]{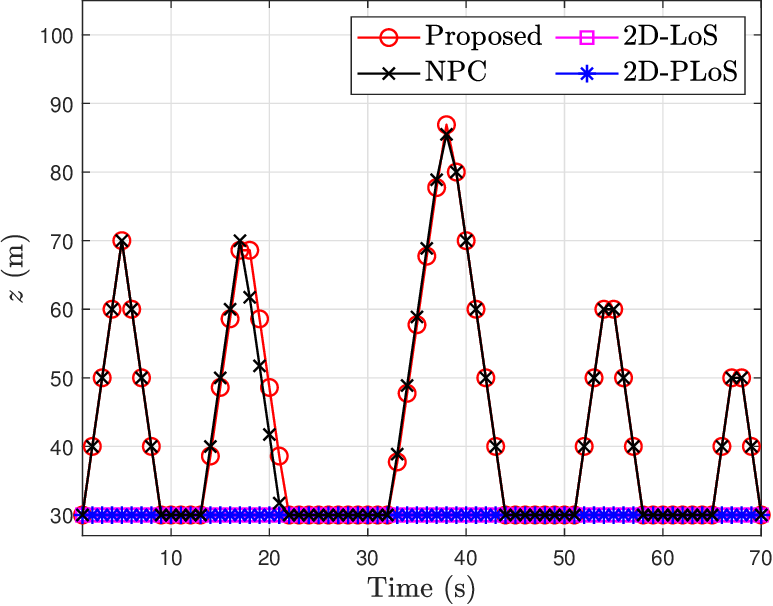}}
	\subfigure[The transmit power of $B$.]{
		\label{fig05c}
		\includegraphics[width = 0.31  \textwidth]{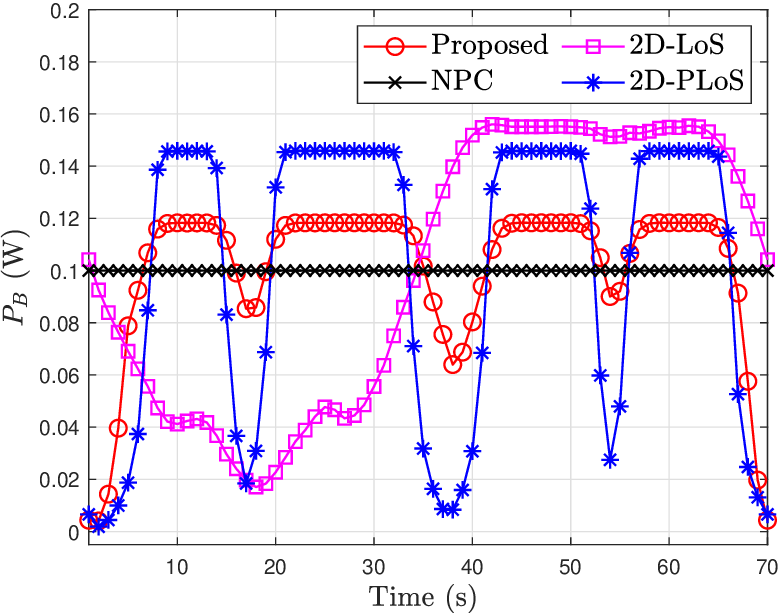}}
	\subfigure[The achievable rate of $B$.]{
		\label{fig05d}
		\includegraphics[width = 0.31  \textwidth]{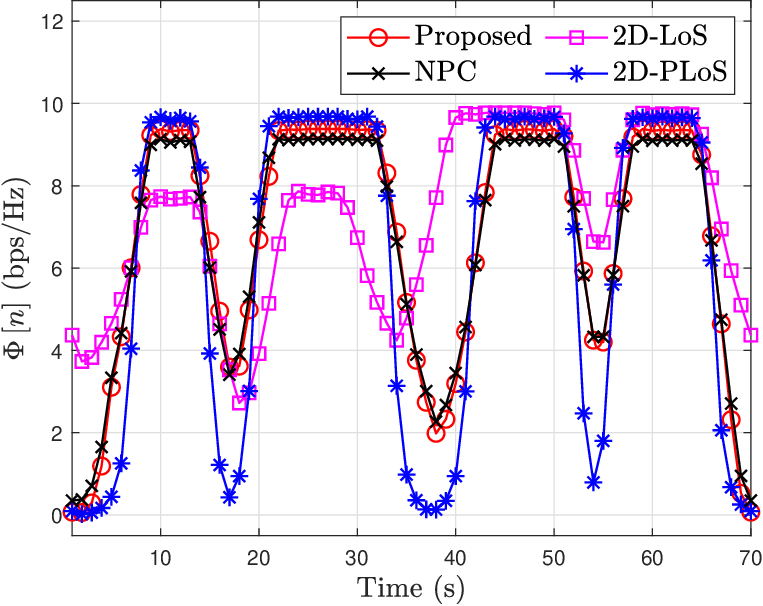}}
	\subfigure[The average rate versus varying $T$ with different schemes.]{
		\label{fig05e}
		\includegraphics[width = 0.31  \textwidth]{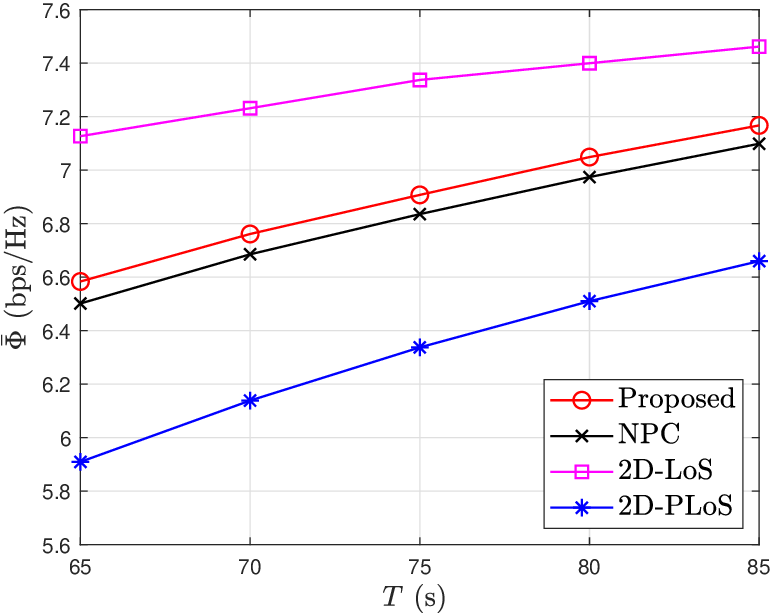}}
	\subfigure[The average rate versus varying $\Gamma$ with different schemes.]{
		\label{fig05f}
		\includegraphics[width = 0.31  \textwidth]{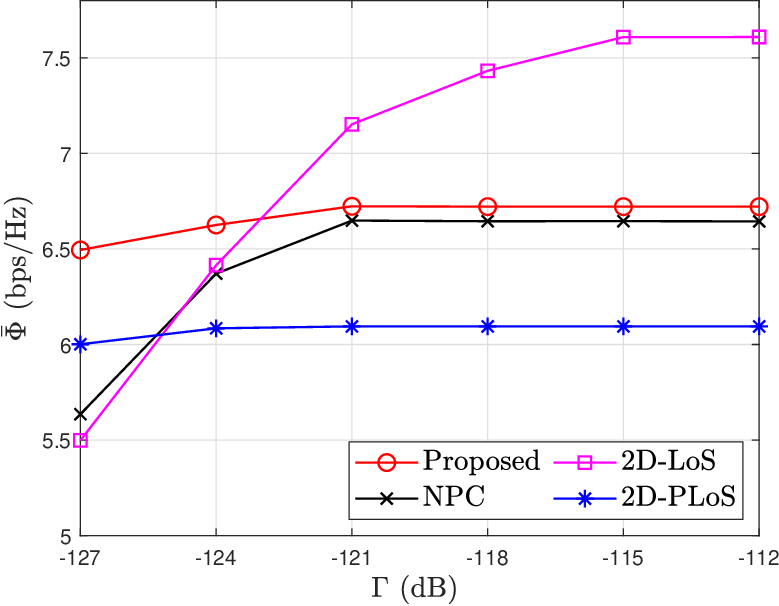}}
	\caption{{Simulation results with different schemes.}}
	\label{fig05}
\end{figure*}
\begin{figure*}[t]
	\centering
	\subfigure[The horizontal trajectory of $B$.]{
		\label{fig06a}
		\includegraphics[width = 0.31  \textwidth]{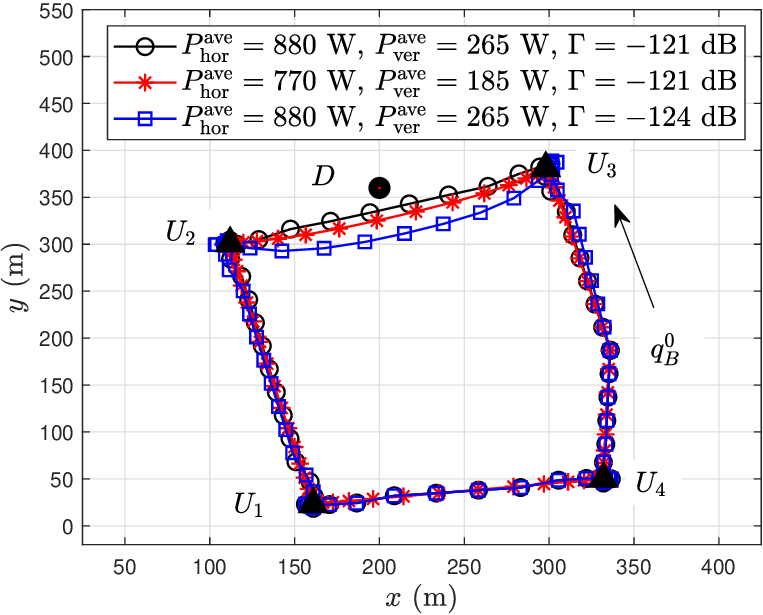}}
	\subfigure[The vertical attitude of $B$.]{
		\label{fig06b}
		\includegraphics[width = 0.31  \textwidth]{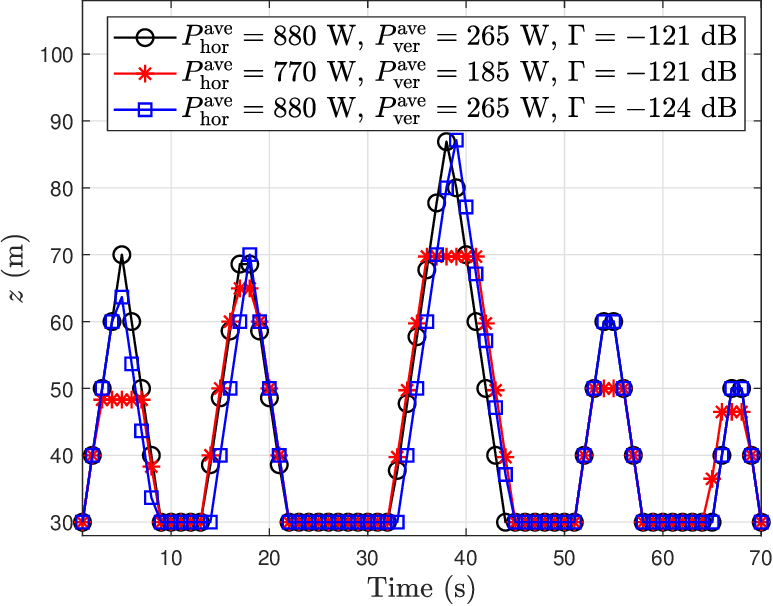}}
	\subfigure[The transmit power of $B$.]{
		\label{fig06c}
		\includegraphics[width = 0.31  \textwidth]{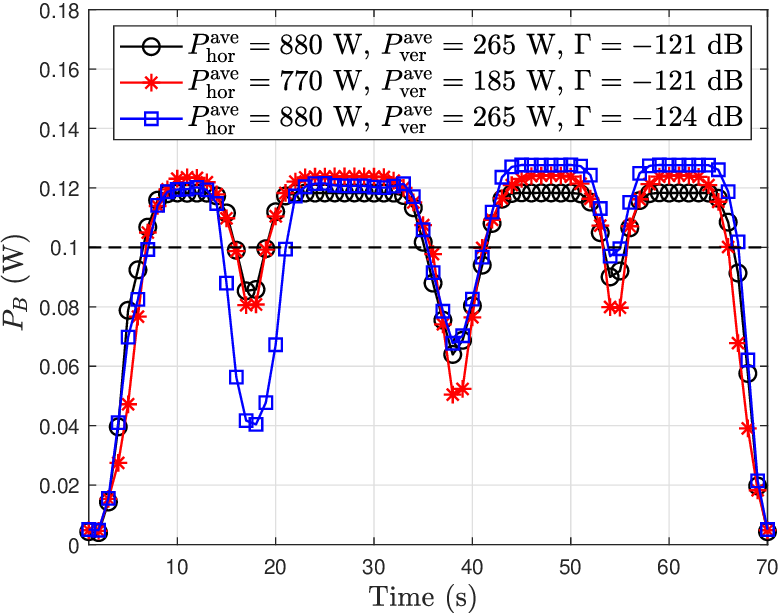}}
	\subfigure[The horizontal speed of $B$.]{
		\label{fig06d}
		\includegraphics[width = 0.31  \textwidth]{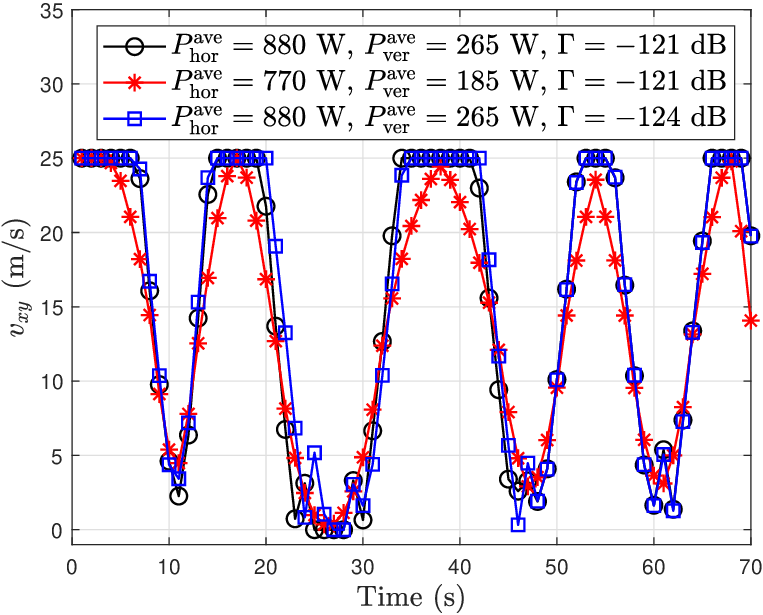}}
	\subfigure[The vertical speed of $B$.]{
		\label{fig06e}
		\includegraphics[width = 0.31  \textwidth]{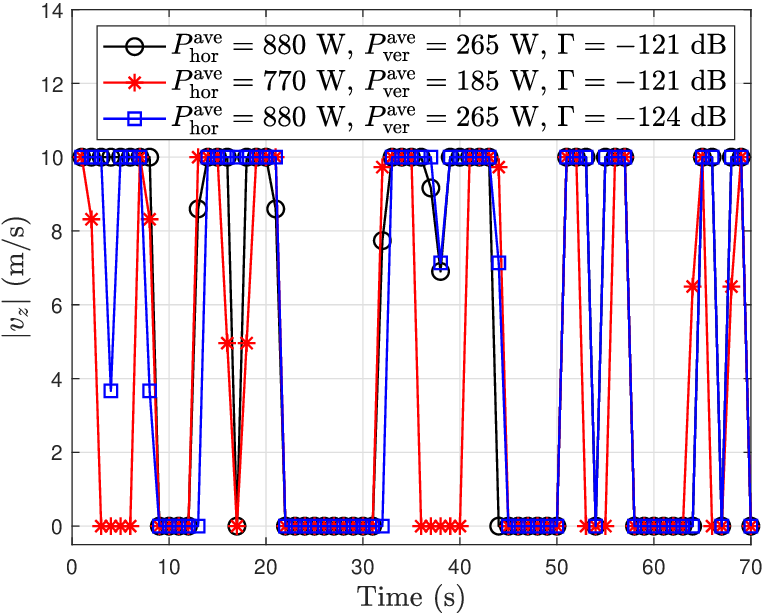}}		
	\subfigure[The achievable rate of $B$.]{
		\label{fig06f}
		\includegraphics[width = 0.31  \textwidth]{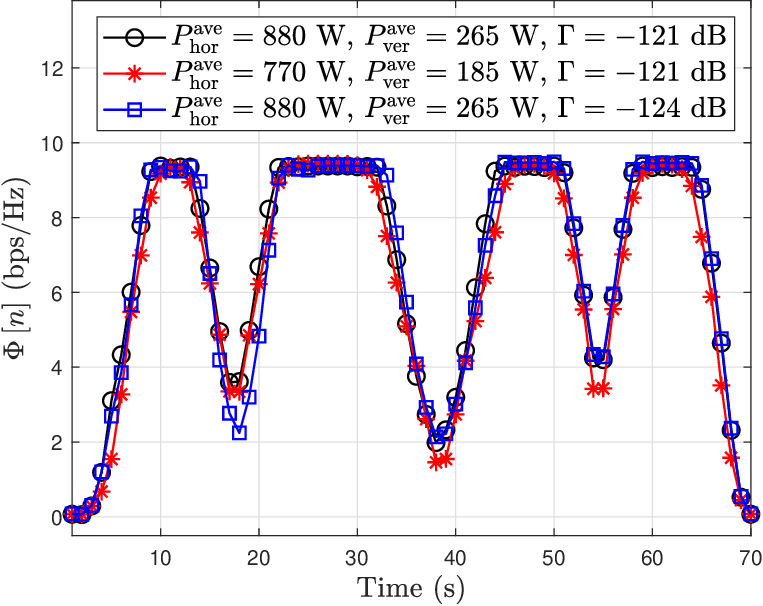}}
	\caption{{Simulation results of the proposed scheme for varying $P_{\lim}$ and $\Gamma$.}}
	\label{fig06}
\end{figure*}

\section{Numerical Results}
\label{Simulation}

In this section, extensive simulations are conducted to validate the effectiveness and the convergence of the proposed algorithm.
To illustrate the advantages of the proposed algorithm, the following three benchmark schemes are compared.
\begin{enumerate}
	\item  Benchmark I: $B$ transmit signals with the average power and both 3D trajectory of $B$ and user scheduling are jointly optimized based on the PLoS model, which is denoted by `NPC'.
	
	\item Benchmark II: The 2D trajectory of $B$, the transmission power of $B$, and user scheduling are jointly optimized based on the LoS model, which is denoted by `2D-LoS'.
	
	\item Benchmark III: $B$ works with fixed vertical trajectory and its horizontal and transmission power of $B$, and user scheduling are jointly optimized based on the PLoS model, which is denoted by `2D-PLoS'.
	
\end{enumerate}

\begin{table}[t]
	\caption{\textit{List of Simulation Parameters.}}
	{
	\begin{center}

		\begin{tabular}{|c|c|}
			\hline
			\textbf{Parameter}   				&\textbf{Value}\\
            \hline
            $\left[ {{{\bf{q}}_B}\left[ 1 \right],{z_B}\left[ 1 \right]} \right]$	& {${\left[ {{\rm{336,187, 30}}} \right]^T}$} \\
            \hline
			${{\mathbf{w}}_{{U_r}}}$             &\makecell[c]{${\left[ {162,23; 112,301;  332,50 ; 298,381} \right]^T}$}\\
			\hline
			${{\mathbf{w}}_{{D}}}$			&\makecell[c]{${\left[ { 200,  360} \right]^T}$}\\
			\hline
			$H_{\min},H_{\max}$  								&30 m, 100 m\\
			\hline
			${\Gamma}$               		&-121 dB\\
			\hline
			${\sigma ^2}$						&-100 dBm\\
			\hline
			$P_B^{\max }$						&0.4 W\\
			\hline
			${P_B^\text {ave }}$			    &0.1 W\\
			\hline
			${\alpha_L},{\alpha_N}$			&2.2, 3.5\\
			\hline
			${a},{b}$			          &11.95, 0.14\\
			\hline
			$V_{\max }$						&25 m/s\\
            \hline
            ${\hat{V}_{\max }}$             &10 m/s \\
            \hline
            $a_{\max }$						&6 ${\rm{m/}}{{\rm{s}}^2}$\\
			\hline
			${\rho _0}$						&-60 dB\\
			\hline
			${\mu}$						&-20 dB\\
			\hline
			${W}$						&100 N\\
			\hline
			${T}$						&70 s\\
			\hline
			${\delta _t}$						&1 s\\
			\hline
			$\varepsilon$						&0.0001\\
			\hline
			${P_{\text {hor }}^{\text {ave }}}$	&880 W\\
			\hline
            ${P_{\text {ver }}^{\text {ave }}}$   & 265 W \\
            \hline

		\end{tabular}
	\end{center}
}
	\label{table3}
\end{table}

Similar to \cite{WuQ2018TWC}, the initial flight trajectory of $B$ is set as a circular trajectory with the center at the geometric center of $U_r$, which is expressed as $\mathbf{C}=\frac{1}{R} \sum\limits_{r=1}^R \mathbf{w}_{U_r}$. To access all CUs as much as possible, the initial flight radius of the UAV is $R_B=\min \left(\frac{V_{\max } T}{2 \pi},\left\|\mathbf{C}-\mathbf{w}_{U_r}\right\|\right)$ \cite{WuQ2018TWC, LeiH2023IoT}. {Details of parameter settings of the UAV communication system are shown in Table \ref{table3} \cite{ZengY2019TWC}, \cite{ZhanC2019WCL}, \cite{DuoB2021ICC}, \cite{YouC2020TWC}, \cite{DuoB2020TVT3D}.} 

Fig.  \ref{fig03} demonstrates the convergence of all the schemes and user scheduling results.
Fig.  \ref{fig03a} indicates that all the schemes converge fast, and the performance with 2D-LoS and 2D-PLOS have the best and the worse performance, respectively. 
Fig.  \ref{fig03b} plots the rate with varying $P_{{\rm{hor}}}^{{\rm{ave}}} $, $P_{{\rm{ver}}}^{{\rm{ave}}} $, and $\Gamma$ and demonstrates that the impact of the power limitation and IT threshold on the average rate of the considered system.
Fig.  \ref{fig03c} indicates that each user is scheduled in turn.

Figs. \ref{fig04a}  and \ref{fig04b} provide the 3D trajectory of $B $ with the different schemes and in the scenarios with different propulsion power limitations, respectively.
From Fig. \ref{fig04a}, it can be observed that compared with the scenarios with the 2D-PLoS scheme, $B$ in the 2D-LoS scheme must keep away from $D$ to meet IT constraints.
This requirement for $ B$ to keep away from $ D$ in the 2D-LoS scheme is due to the exaggerated probability of LoS for the A2G link.
The results in Fig. \ref{fig04b} show that different propulsion power limitation results in different vertical trajectories.

Fig.  \ref{fig05} demonstrates the simulation results under different schemes with $P_{{\rm{hor}}}^{{\rm{ave}}} = 880$ W, $P_{{\rm{ver}}}^{{\rm{ave}}} = 265$ W, and $\Gamma = -121$ dB.
Figs.  \ref{fig05a}, \ref{fig05b}, \ref{fig05c}, and \ref{fig05d} plot the horizontal trajectory, vertical trajectory, transmit power, and instantaneous achievable rate of different schemes, respectively.
It can be observed from Fig. \ref{fig05a} that different schemes achieve the same horizontal trajectory in the region away from $D$. However, in the area near $D$, the results of the proposed scheme, NPC, and 2D-PLoS are different from those of the 2D-LoS scheme.
This is because the A2G link is treated as LoS in the 2D-LoS scheme, $B$ must move horizontally away from $D$ to satisfy the IT constraint.
As can be seen in Fig. \ref{fig05b}, when $B$ moves away from the TNs, it first raises its altitude and then dives down to get a better elevation angle, which also can be found from Figs. \ref{fig04a} and \ref{fig04b} and stated in \cite{LeiH2024TCOM}.
Fig.  \ref{fig05c} shows the the transmit power is optimized to meet the IT constraint and maximize performance. 
In particular, in the 2D-PLoS scheme, the higher power is used when the UAV is near TNs, while the lower power is used when the UAV is far away from TNs. 
That is, performance is maximized by allocating different power at different locations. 
Moreover, in the 2D-LoS scheme, since A2G links are treated as LoS, the transmit power is constrained by the IT constraint.
Specifically, the transmit power in 2D-LoS decreases as the distance between $B$ and PU increases.  
As can be seen from Fig. \ref{fig05d}, the considered system with the 2D-LoS has the lowest performance in the areas near $D$ because the lowest transmit power is used subject to the IT constraints. 
The proposed scheme's performance in the regions near the TNs is between that of the NPC and the 2D-PLoS schemes because both the transimit power and the vertical trajectory are optimized to obtain a larger elevation angle (the larger the probability of LoS).
Figs. \ref{fig05e} and \ref{fig05f} show the relationship between the average rate and varying $T$ and IT thresholds, respectively.
From Fig. \ref{fig05e}, one can observe that, same as in Fig. \ref{fig03a}, the performance improves as $T$ increases because $B$ can spend more time hovering over TNs.
The 2D-LoS scheme has the best performance. The reason is, in the areas away from PU, more power can be used to maximize the achievable rate. 
In addition, in the 2D-LoS scheme, the A2G link is always considered LoS $\left(P_{BX}^{\mathrm{L}} = 1\right)$.
Fig. \ref{fig05f} demonstrates the effect of the IT thresholds on the average rate with different schemes. 
One can observe that the average rate increases as the IT threshold becomes more relaxed. 
This is because, with the higher IT threshold, $B$ can use higher transmit power and thus obtain higher performance. When $\Gamma$ increases to a specific range, the system enters the no-cognitive state, and the system performance is independent of the IT threshold. 
In the low-$\Gamma$ region, the requirement for the CUs is very harsh and the transmit power is limited to the IT constraint. 
Specifically, the NPC scheme does not converge when $\Gamma = -128$ dB.
Limited by the IT constraint, the performance of the 2D-LoS scheme is the worst in the low-$\Gamma$ region. 
However, in the high-$\Gamma$ region, the system enters the no-cognitive state and higher transmission power can be used, thus, the 2D-LoS has the best performance.

Fig.  \ref{fig06} describes the simulation results of the proposed scheme with varying $P_{{\rm{hor}}}^{{\rm{ave}}} $, $P_{{\rm{ver}}}^{{\rm{ave}}} $, and $\Gamma$.
Fig.  \ref{fig06a} show that, in the scenarios with lower propulsion power and lower IT thresholds, $B$ is needed away from $D$ in the region near $D$ to meet the IT constraint. 
This is because the lower propulsion power responds to the lower horizontal speed, shown at Fig.  \ref{fig06d}, and the lower IT threshold denotes that the requirements for the cognitive users are more strict, so $B$ needs to be away from $D$.
With the same reason, in the area close to $D$, the elevation angle cannot reach the maximum to ensure that IT constraints are met, presented in Fig.  \ref{fig06b}.
In areas away from the PU, IT constraints are not limited and vertical trajectory is maximized.
Fig.  \ref{fig06c} shows that in the scenarios with small $\Gamma$, the UAV uses lower power in the area near the PU. In areas away from PU, higher power is used.
Figs.  \ref{fig06d} and \ref{fig06e} plot the optimized horizontal and vertical speeds for the different power and IT thresholds, respectively.
It can be observed that the larger propulsion power budget, the longer hovering time and the longer fall time, which result in higher performance.

\section{Conclusion}
\label{Conclusions}

This work investigated the achievable rate of the underlay IoT system with an energy-constrained UAV under PLoS channels.
The achievable rate of the considered systems was maximized by jointly considering the UAV's 3D trajectory, transmission power, and user scheduling, which is a nonlinear mixed-integer non-convex problem.
The lower bound of the average achievable rate was utilized and the original non-convex problem was transformed into several solvable convex subproblems by using BCD and SCA techniques, and an efficient iterative algorithm was proposed.
The numerical results not only verified the convergence and effectiveness of the algorithm but also illustrated the impact of propulsion power and interference thresholds on the average achievable rate.

\begin{appendices}
	\section{Proof of Lemma 1  }	
	\label{lemma1}

For simplicity, we define a bivariate function $f(x, y) =\frac{1}{x} \log_2 \left(1+\frac{A}{y}\right)$, where $A \geq 0$, $x$ and $y$ are both positive variables of the function. The Jacobian of $f(x, y)$ is
\begin{align}
		\nabla f(x, y) & =\left[\begin{array}{ll}
			\frac{\partial f(x, y)}{\partial x} & \frac{\partial f(x, y)}{\partial y}
		\end{array}\right] \nonumber\\
		& =\left[\begin{array}{ll}
			-\frac{1}{x^2}  \log_2(1+\frac{A}{y}) & -\frac{A\log_2e}{xy^2(1+\frac{A}{y})}
		\end{array}\right].
		\label{H232}
\end{align}

From the above Jacobian matrix, we can obtain the Hessian matrix of $f(x, y)$ as
\begin{align}\label{hessin}
		\nabla^2 f(x, y) &=\left[\begin{array}{ll}
			\frac{\partial^2 f(x, y)}{\partial x^2} & \frac{\partial^2 f(x, y)}{\partial x \partial y} \\
			\frac{\partial^2 f(x, y)}{\partial y \partial x} & \frac{\partial^2 f(x, y)}{\partial y^2}
		\end{array}\right] \nonumber\\
		&=\left[\begin{array}{cc}
			\frac{2}{x^3}  \log_2(1+\frac{A}{y}) &  \frac{A\log_2e}{x^2y^2(1+\frac{A}{y})} \\
			\frac{A\log_2e}{x^2y^2(1+\frac{A}{y})} & \frac{A(2y+A)\log_2e}{xy^4(1+\frac{A}{y})^2}
		\end{array}\right] .
\end{align}

We need to scale $\log_2(1+\frac{1}{x})$ to take its lower bound. When $x\geq 0$, we can use Lagrange's mean value theorem to determine $\log_2(1+\frac{1}{x})$$\geq$$\frac{\log_2e}{1+x}$. Therefore, we replace it with the lower bound of $\log_2(1+\frac{A}{y})$ and find the determinant of its Hessian matrix. Furthermore, the determinant of the Hessian matrix of $f(x, y)$ is
\begin{align}
		\left|\nabla^2 f(x, y)\right|&=  \frac{2}{x^4y^4}  \frac{A(2y+A)\log_2e}{(1+\frac{A}{y})^2} \log_2 \left(1+\frac{A}{y}\right)  \nonumber\\
		& -\frac{1}{x^4y^4}  \frac{A^2(\log_2e)^2}{(1+\frac{A}{y})^2} \nonumber\\
		&\geq\frac{1}{x^4y^4}  \frac{A^2(2y+A)(\log_2e)^2}{(1+\frac{A}{y})^2(y+A)} \nonumber\\
		& -\frac{1}{x^4y^4}  \frac{A^2(\log_2e)^2}{(1+\frac{A}{y})^2} \nonumber\\
		&=\frac{1}{x^4y^4}  \frac{A^2y(\log_2e)^2}{(1+\frac{A}{y})^2(y+A)} \geq 0.
		\label{tuidao}
\end{align}

According to (\ref{hessin}) and (\ref{tuidao}), for any given $A\geq0$, We can determine that the determinant of the cofactor matrix of the Hessian matrix of function $f (x, y) $ is greater than or equal to 0. Therefore, the Hessian matrix of $f(x, y)$ is a semi-positive definite matrix, so $f(x, y)$ is a convex function. 

\end{appendices}

\end{document}